\documentclass[aps,prd,eqsecnum, showpacs]{revtex4}
\usepackage{graphics}
\usepackage{psfrag}
\begin{document}
\title{Gravitational quasinormal modes for Kerr Anti-de Sitter black holes}
\author{M. Giammatteo and Ian G. Moss}
\affiliation{School of Mathematics and Statistics,\\University of Newcastle
upon Tyne, NE1
7RU U.K.}
\date{ December 2004}

\begin{abstract}
We investigate the quasinormal modes for gravitational perturbations of
rotating black holes in four dimensional Anti-de Sitter (AdS) spacetime. The
study of the quasinormal frequencies related to these modes is relevant to the
AdS/CFT correspondence. Although results have been obtained for Schwarzschild
and Reissner-Nordstrom AdS black holes, quasinormal frequencies of Kerr-AdS
black holes are computed for the first time. We solve the Teukolsky equations
in AdS spacetime, providing a second order and a Pade approximation for the
angular eigenvalues  associated to the Teukolsky angular equation. The
transformation theory and the Regge-Wheeler-Zerilli equations for Kerr-AdS are
obtained.
\end{abstract}
\maketitle

\section{Introduction}
The holographic principle in field theory states that, in the presence of
gravity, physics in a region of space-time can be encoded on a
lower-dimensional surface. The Maldacena $AdS/CFT$ conjecture \cite{maldacena}
provides explicit examples for some of these holographic encodings.
For example, IIB string theory on the background $AdS_5 \times S^5$ or
$M-$theory on the background $AdS_4\times S^7$ are claimed to be
mathematically equivalent
(dual) to certain quantum field theories on the boundary at infinity of the
relevant $AdS$ space. The isometry groups of the $AdS$ spaces become conformal
groups acting on the boundary quantum theories. In the case of $AdS_{5}\times
S^{5}$, the quantum field theory is $\mathcal{N}=4$ Super Yang-Mills gauge
theory with gauge group $U(N)$ and coupling constant $g_{YM}$. This particular
Yang-Mills theory is also a conformal field theory $(CFT)$.

More specifically, if one introduces a generating function for correlation
functions of operators $O_j$ in the Yang-Mills theory, then the $AdS/CFT$
correspondence states that \cite{gubser, witten},
\begin{equation}
\left<
e^{-\int \phi^{(0)}_{j} O_j(x)d^{D}x}
\right>_{CFT}=
\mathcal{Z}_{string}[\phi_k].
\end{equation}
In this equation, on the left-hand side, we have the generating function with
operators integrated against arbitrary density functions, $\phi^{(0)}_{j}$, on
the boundary of $AdS$. The boundary has dimension $D= 4, 3$ according to the
version of the correspondence we consider. The function 
$\mathcal{Z}_{string}[\phi_k]$
is the string partition function with string coupling $g_s$, evaluated in the
$AdS$ background. The fields, $\phi_{k}$, satisfy the boundary conditions of
approaching $\phi^{(0)}_{j}$ at infinity.

A limit can be taken to reduce the string theory to its supergravity limit.
This may be done by a combination of taking the string tension to infinity
$(\alpha'\rightarrow 0)$ and sending $g_{s}N\rightarrow\infty$. A second limit
is obtained by making the approximation $(g_{s}\rightarrow 0)$ of the
supergravity partition function to $\exp(-S)$, where the supergravity action
$S[\phi_k]$ is the action of the classical supergavity solution that satisfies
the boundary conditions mentioned above. Although the strongest form of the
correspondence is the string theory expression, in practice it is much easier
to work with supergravity actions, and it is for this reason that most results
have been obtained in this context. On the quantum field theory side of the
correspondence, this approximation amounts to taking both $N$ and
$g^{2}_{YM}N$ large. The classical supergravity action should holographically
capture the large-$N$ limit of strongly coupled quantum Yang-Mills theory.

A quite large amount of work has been produced in the last few years on the
$AdS/CFT$ conjecture and an extensive introduction can be found in
\cite{aharony, petersen}. In the $AdS/CFT$ correspondence, the study of $AdS$
black holes has a direct interpretation in terms of the dual conformal field
theory on its boundary. The duality predicts that the retarded $CFT$
correlation functions are in one to one correspondence with Green's functions
on Anti-de Sitter space with
appropriate boundary conditions \cite{danielsson,kalyana,birmingham,son}. 
Furthermore, as mentioned in \cite{horowitz}, it is assumed that a large static
black hole in $AdS$ space corresponds to a thermal state in the $CFT$ on the
boundary. Perturbing the black hole with a scalar field, for instance, is
equivalent to perturbing this thermal state. The perturbed system is expected,
at late times, to approach equilibrium exponentially with a characteristic
time-scale. This time-scale is inversely proportional to the imaginary part of
the poles of the correlators of the perturbation operator. It seems that these
relaxation time-scales are quite complicated to calculate in the $CFT$,
therefore their computation is conveniently replaced by the evaluation of the
quasinormal frequencies in the $AdS$ bulk space. 

In \cite{lemos} and \cite{moss}, the authors went beyond the scalar
perturbation treated in \cite{horowitz} and they considered electromagnetic
and gravitational perturbations.  In one of the most recent works on the
subject, Berti and Kokkotas \cite{berti} confirm and extend previous results
on scalar, electromagnetic and
gravitational perturbations of static $AdS$ black holes and analyse, for the
first time, Reissner-Nordstrom Anti-de Sitter black holes and calculate their
quasinormal frequencies. Naturally, perturbations associated with other kinds
of fields than the scalars treated in \cite{horowitz} will decay at different
rates given by their quasinormal mode frequencies. 

Quasinormal modes have recently become relevant to quantum gravity for a
different reason. This arises from Hod's proposal to
describe quantum properties of black holes from their classical oscillation
spectrum \cite{hod}. The quest for a consistent quantization of black holes
was initiated many years ago by Bekenstein \cite{bekenstein1, bekenstein2}
who, based on semi-classical arguments, conjectured that the horizon area of a
(non extremal) black hole should have a discrete eigenvalue spectrum. The
eigenvalues of this spectrum were expected to be uniformly spaced. Hod noted
that the real parts of the asymptotic quasinormal frequencies of a static
black hole of mass $M$, as computed by Nollert \cite{nollert}, can be written
as $\omega_R=T ln 3$, where $T$ is the black hole Hawking's temperature.
He then exploited Bohr's correspondence principle, asserting that 'transition
frequencies at large quantum numbers should equal classical oscillation
frequencies', to conjecture that variations in the black hole mass due to
quantum processes should be given by $\Delta M=\hbar\omega_R$. Finally, using
the first law of black hole thermodynamics and the equivalence between entropy
and the surface area, he deduced the equal spacing in the area spectrum of a
static hole.

Hod's conjecture has brought further motivations in the field of black hole
physics and has provoked several attempts to evaluate highly damped
quasinormal frequencies of Kerr black holes. It is, in fact, straightforward
to ask whether this conjecture is valid for more general black holes than the
static ones; if asymptotic frequencies depend in these more general cases on
black hole charge and/or angular momentum; if the presence of a cosmological
constant should imply a modification of the proposal. At the moment these
ideas of universality in Hod's proposal have not been confirmed \cite{natario}. 

A brief outline of this paper is the following. In section \ref{two} we review
some of the main features of $AdS$ spacetime and its static and rotating black
hole solutions. In section \ref{three} we solve the radial Teukolsky equation.
Then, we concentrate on the angular equation and propose both a second order
and a Pade approximation of the associated eigenvalues. In section IV we
introduce transformation theory and obtain the Regge-Wheeler equation. These
results are exploited for the explicit determination of the quasinormal
frequencies and some considerations on their asymptotic spacing. All the
numerical results are reported in section \ref{numres}. We conclude by a final
discussion of the results we have obtained and with some ideas for future
work. In the appendix we give the explicit recurrence relation for the series
solution of the radial Teukolsky equation.

\section{AdS spacetime and its rotating black hole solutions}
\label{two}
Rotating black holes in four dimensions with a cosmological constant were
first constructed by Carter \cite{carter} and Demianski \cite{demianski}. In
the asymptotically de Sitter case, the black hole solutions have cosmological
horizons. Thermal equilibrium is only possible when the black hole horizon has
the same temperature as the de Sitter horizon\cite{mellor}. The asymptotically
Anti-de Sitter case has no cosmological horizons and the black hole, even if it
is rotating, can be in equilibrium with radiation out to the conformal
boundary at infinity \cite{hawking,caldarelli,gibbons04}.

Anti-de Sitter space and the associated stationary black hole
solutions can be described by a metric with radial component $g_{rr}$ of the
form
\begin{equation}
{r^2+a^2 \cos^{2}\theta\over \Delta_r}.
\end{equation}
The polynomial $\Delta_r(r)$ vanishes at the event horizon, which we shall
denote by $r_+$.  The event horizon is also a Killing horizon, that is it is a
surface on which a Killing field $\xi^{a}$ become null. 
The surface gravity $\kappa$ is defined in terms of $\xi^{a}$ by,
\begin{equation}
\nabla^{a}(\xi^{b}\xi_{b})=-2\kappa\xi^{a}.
\end{equation}
For a stationary and axisymmetric black hole with timelike and axial killing
fields $\partial_t$ and $\partial_\phi$, the killing vector field $\xi^{a}$ is
of the form
\begin{equation}
\xi=\partial_t+\Omega_H\partial_\phi,
\end{equation}
The constant $\Omega_H$ is the angular velocity of the horizon. 

An explicit expression for the surface gravity of the event horizon is given by
\cite{gibbons},
\begin{equation}\label{c4e0}
\kappa=\frac{\Delta'(r_+)}{2(a^2+r_+^2)}.
\end{equation}
The Hawking temperature of the black hole is related to the surface gravity
through the famous result,
\begin{equation}
T=\frac{\kappa}{2\pi}.
\end{equation}
The version of the $AdS/CFT$ conjecture which concerns us here is the
correspondence between $M-$theory on $AdS_4 \times S^7$ and
supersymmetric conformal Yang-Mills field theory on the conformal boundary at
infinity. Crucial to the $AdS/CFT$ conjecture is the fact that the $AdS$ space
and its related black hole solutions are conformal to a rotating version of
the Einstein static Universe at infinity. The angular velocity of the space at
infinity is denoted by $\Omega$. The combination 
$(T,\Omega)$ form a set of thermodynamic variables 
\cite{hawking,caldarelli,gibbons04}.

\begin{figure}[ht!]
\begin{center}
 \leavevmode
 \scalebox{0.5}{\includegraphics{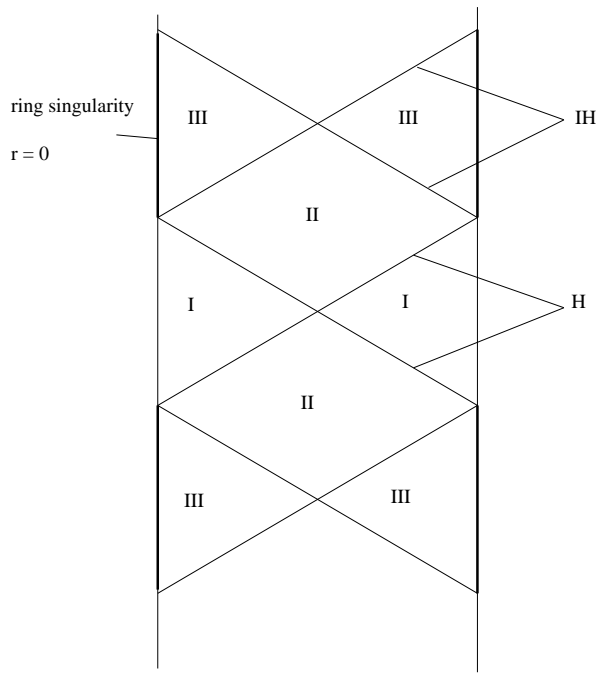}}
\caption{The Penrose diagram of a Kerr black hole in Anti-de Sitter space-time
is shown. $IH$ and $H$ are the inner and event horizon of the black hole.
Regions $I$, bound by vertical segments $(r=\infty)$, represent the exterior
space. Regions $II$ are the regions between the inner $(r=r_-)$ and the event
horizon of the black hole. Regions $III$ correspond to $r<r_-$. The bold
vertical segments represent the ring singularity at $r=0$ in the equatorial
plane.}
\label{PenroseRads}
\end{center}
\end{figure}

For Kerr-AdS spacetime with cosmological constant $\Lambda$,
\begin{equation}
\Delta_r=(a^2+r^2)(1-\frac{1}{3}\Lambda r^2)-2Mr
\end{equation}
In Fig. \ref{PenroseRads}, we show the conformal compactification for the
covering space of Kerr-$AdS$ spacetime. The conformal boundary is the boundary
of region I.

The black hole event horizon is the largest real root of $\Delta_r=0$,
\begin{equation}\label{c4e2}
(r^2+a^2)(1+\alpha^2 r^2)-2Mr=0,
\end{equation}
where $\alpha^2=-\Lambda/3$ is the inverse of the Anti-de Sitter radius. From
Eq. \ref{c4e2} and the definition of the surface gravity \ref{c4e0}, we derive
an expression for the temperature $T$,
\begin{equation}\label{c4e3}
T={3\alpha^2 r_+^4+(1+\alpha^2 a^2)r_+^2-a^2\over 4\pi r_+(r_+^2+a^2)}.
\end{equation}
The rotation of the asymptotic space is given by
\cite{hawking,caldarelli,gibbons04}
\begin{equation}
\Omega={a(1+\alpha^2r_+^2)\over a^2+r_+^2}
\end{equation}

\begin{figure}[ht!]
\begin{center}
\leavevmode
\includegraphics{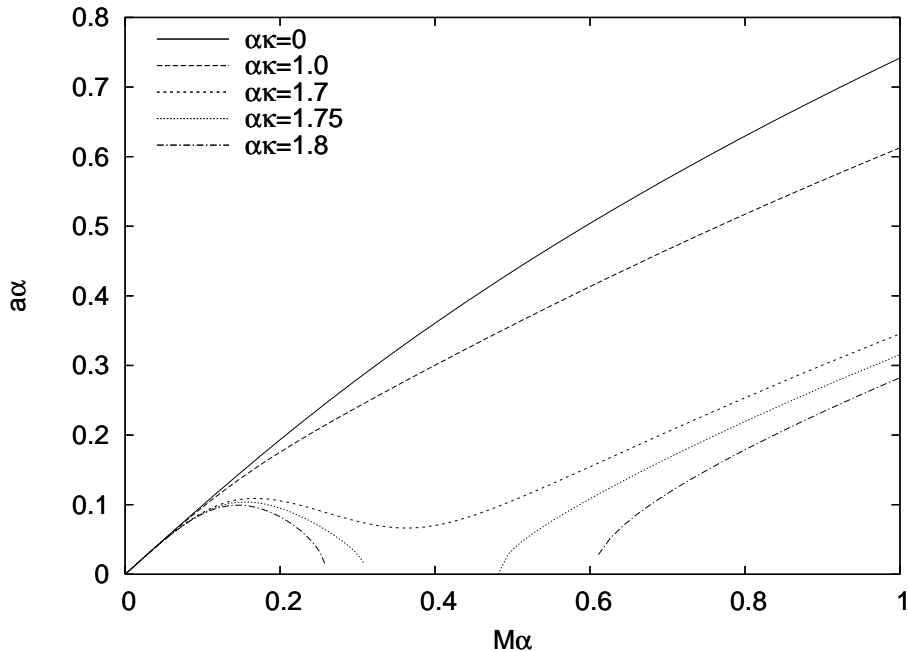}
\caption{Isotherms related to Kerr-$AdS$ black holes are plotted in the
$a\alpha$ and $M\alpha$ parameter space.}
\label{C4F4}
\end{center}
\end{figure}

Fig. \ref{C4F4} shows contours of constant $\kappa$ in the $a\alpha$ and
$M\alpha$ parameter space. The $\kappa=0$ solutions are extremal black holes
and above this line in the diagram the solutions become naked singularities.
Fixing the value of the angular parameter $a$ and $\kappa$ we have unique
masses and therefore unique solutions for $a\alpha\geq 0.1$, but up to $3$
different solutions for $a\alpha<0.1$.

\begin{figure}[ht!]
\begin{center}
\leavevmode
\includegraphics{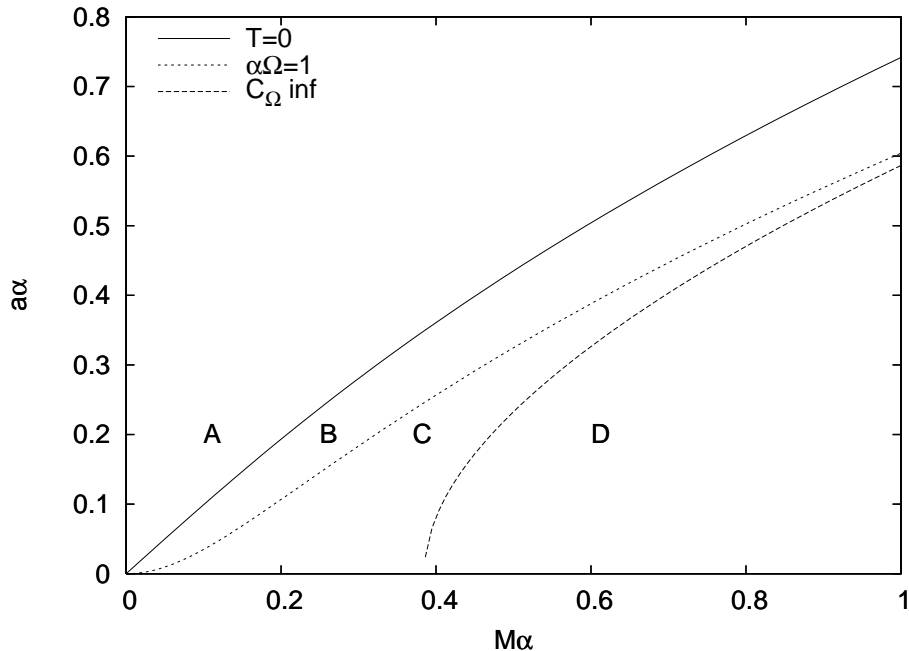}
\caption{This figure of the $a\alpha$ and $M\alpha$ parameter space shows
various stability limits, $\alpha\Omega=1$ and the critical line where the
thermal capacity $C_\Omega$ diverges. The parameter space is divided into
regions:  $A$ with only naked singularities;  $B$ where $\alpha\Omega>1$ and
$C_\Omega<0$; $C$ where $\alpha\Omega<1$ and $C_\Omega<0$ and $D$ where
$\alpha\Omega<1$ and $C_\Omega>0$.}
\label{C4F5}
\end{center}
\end{figure}

The transformation from the metic parameters $M$ and $a$ to the thermodynamic
variables becomes singular where the
Jacobian
\begin{equation}
{\partial(T,\Omega)\over\partial(r_+,a)}=
{r_+^2(3\alpha^2r_+^2-1)-a^2(1+\alpha^2r_+^2)\over 4\pi (r_+^2+a^2)^2}
\end{equation}
vanishes. Vanishing of the Jacobian also corresponds to the divergence of the
heat capacity at constant $\Omega$. This critical line is plotted in Fig.
\ref{C4F5}. Another important feature of the parameter space is set by the
condition $\alpha\Omega<1$. If $\alpha\Omega>1$, then the killing field $\xi$
becomes non-timelike beyond a certain radius and the black hole
radiation can not remain in thermal equilibrium. This allows us to divide the
paramter space into the four regions $A, B, C,D$. Region $A$ is the region with
no event horizons. Regions $B$ and $C$ both have negative specific heat.
Region $D$ is the region of positive specific heat and $\alpha\Omega<1$.

\section{Quasinormal frequencies of Kerr AdS black holes}
\label{three}
 The full Kerr $AdS$ metric can be found in \cite{mellor}, for instance, and it
can be expressed as
\begin{equation}
ds^2  =  -\frac{\Delta_r}{(r^2+\mu^2)}\omega_1^2+
\frac{\Delta_{\mu}}{(r^2+\mu^2)}\omega_2^2
+\frac{r^2+\mu^2}{\Delta_{r}}dr^2+
\frac{r^2+\mu^2}{\Delta_{\mu}}d\mu^{2}
\end{equation}
where $r, \theta$ and $\phi$ are spherical coordinates, with $\mu=a\cos\theta$.
The forms $\omega_1$ and $\omega_2$ are
\begin{equation}
\omega_1=dt-a\chi^{-2}\sin^2\theta\,d\phi,\qquad
\omega_2=dt-a^{-1}\chi^{-2}(r^2+a^2)\,d\phi
\end{equation}
and the metric functions are given by
\begin{eqnarray}
\Delta_{r}&=&(a^2+r^2)(1-\frac{1}{3}\Lambda r^2)-2Mr ,\\
\Delta_{\mu}&=&(a^2-\mu^{2})(1+\frac{1}{3}\Lambda \mu^{2}) ,\\
\chi^{2}&=& 1+\frac{1}{3}\Lambda a^2.
\end{eqnarray}

We shall make use later of the radial tortoise coordinate, defined by
\begin{equation}
\frac{d}{dr^{*}}=\frac{\Delta_r}{(a^2+r^{2})}\frac{d}{dr}.
\end{equation}
Using the the definition of surface gravity given in Eq. \ref{c4e0}, applied to
each of the roots of $\Delta_r=0$ (two of them complex), we can express
$r^{*}$ in the form
\begin{equation}
r^{*}=\frac{1}{2\kappa_+}\ln\alpha(r-r_+)+
\frac{1}{2\kappa_-}\ln\alpha(r-r_-)+
\frac{1}{2\kappa_3}\ln\alpha(r-r_3)+
\frac{1}{2\bar{\kappa}_3}\ln\alpha(r-\bar r_3),
\end{equation}
Note that $r^{*}\in (-\infty, 0)$, with $r^{*}\rightarrow 0$ as
$r\rightarrow\infty$ as a consequence of the identity
$\kappa_+^{-1}+\kappa_-^{-1}+\kappa_3^{-1}+\bar\kappa_3^{-1}=0$. (In the
ordinary Kerr solution, $r_*\to \infty$ as $r\to\infty$).

The generalisation of Teukolsky's equations and their separation for
asymptotically de-Sitter and Anti de-Sitter spacetimes, was first derived by
Chambers and Moss \cite{chambers}. They looked for separable solutions for the
Weyl scalar of the
form
\begin{equation}
\Psi_0= e^{+i\omega t}e^{im\phi}R(r)S_{l}(\theta),
\end{equation}
where $S_{l}(\theta)$ can be found in \cite{chandra}, and $l$ is the angular
mode number characteristic of the spherical harmonics. (In \cite{chambers}, the
time coordinate was scaled differently and the frequency was a factor of
$\chi^2$ larger than the frequency which we are using here.) In what follows
we will drop the index $l$ when we talk about the angular function $S_l$.

Following \cite{chambers}, we introduce new radial and angular
derivatives 
\begin{eqnarray}
D_{n}&=& \partial_{r}+\frac{iK_{r}}{\Delta_{r}}+
\frac{n\Delta^{'}_{r}}{\Delta_{r}} ,
\hspace{0.8cm}D^{\dagger}_{n}=\partial_{r}-
\frac{iK_{r}}{\Delta_{r}}+\frac{n\Delta^{'}_{r}}{\Delta_{r}},\\
L_{n}&=&\partial_{\mu}+\frac{K_{\mu}}{\Delta_{\mu}}+
\frac{n\Delta^{'}_{\mu}}{\Delta_{\mu}} 
\hspace{0.8cm}L^{\dagger}_{n}=\partial_{\mu}-
\frac{K_{\mu}}{\Delta_{\mu}}+
\frac{n\Delta^{'}_{\mu}}{\Delta_{\mu}},\label{c4e19}
\end{eqnarray}
with $K_{r}$ and $K_{\mu}$ defined as
\begin{eqnarray}
K_{r}       &=&\chi^{2}am+(a^2+r^2)\omega,\\
K_{\mu}     &=&\chi^{2}am+(a^2-\mu^2)\omega.
\end{eqnarray}
The angular and radial functions $S(\theta)$ and $R(r)$, satisfy the two second
order differential equations
\begin{eqnarray}
( L_{-1}\Delta_{\mu} L^{\dagger}_{1}+6\omega\mu-2 \Lambda\mu^{2})S = -
\lambda S \label{c4e4} \\
( D_{-1}\Delta_{r} D^{\dagger}_{1}-6i\omega r-2 \Lambda r^2)R = \lambda
R \label{c4e5}
\end{eqnarray}
where $\lambda$ is a separatin constant. In the non-rotating case,
$\lambda=(l-1)(l+2)$ for $l=2,3,\dots$.

\subsection{Radial equation}
Let us start from the radial differential equation \ref{c4e5}, which can be
re-written as
\begin{equation}
\left\{\left(\partial_{r}+\frac{iK_{r}}{\Delta_r}-\frac{\Delta'_r}
{\Delta_r}+\frac{1}{r}\right)\frac{\Delta_r}
{r^2}\left(\partial_{r}-\frac{iK_{r}}{\Delta_r}+
\frac{\Delta'_r}{\Delta_r}-\frac{1}{r}\right)+
\frac{-6i\omega}{r}
-2\Lambda-\frac{\lambda}{r^2}\right\}R=0.
\end{equation}
This equation has four regular singular points at the four roots of $\Delta_r$;
one at the black hole outer horizon, $r_+$, which is the largest root of
$\Delta_r$, one at the black hole inner horizon, $r_-$, and other two complex
conjugate roots, $r_3$ and $\bar r_{3}$, which are unphysical roots. This
allows us to write $\Delta_{r}$ as 
\begin{equation}
\Delta_r= 
\alpha
\left(r-r_+\right)\left(r-r_-\right)\left(r-r_{3}\right)
\left(r-\bar r_{3}\right).
\end{equation}
where $\alpha^{2}=-\Lambda/3$ is the inverse of the Anti-de Sitter radius.
Next, we set $x=r^{-1},$ so that we can write 
\begin{equation}
\Delta_r=p x^{-4},
\end{equation}
with
\begin{equation}
p=\alpha^{2}
\left(x-x_{1}\right)
\left(x-x_{2}\right)\left(x-x_{3}\right)\left(x-\bar x_{3}\right).
\end{equation} 
Note that $x_{1}, x_{2}, x_{3}, \bar{x}_{3}$ are just the inverse of the roots
of $\Delta_{r}$ in terms of the new variable $x$.

Using the same coordinate inversion we can rewrite $K_{r}$ as
\begin{equation}
K_{r} = -x^{-2} q,
\end{equation} 
with 
\begin{equation}\label{C425}
q = - \chi^2amx^2+\left(1+a^2 x^2\right)\omega.
\end{equation}
It is also useful, for future reference, to re-write $p$ as 
\begin{equation}\label{C426}
p=\alpha+\left(1+a^2 \alpha^2\right)x^2-2Mx^3+a^2 x^4,
\end{equation}
 In the new variable $x$ the radial equation becomes
\begin{equation}\label{c4e6}
\left\{\left(\partial_{x}+\frac{iq}{p}-\frac{p'}{p}+\frac{3}
{x}\right)p\left(\partial_{x}-\frac{iq}{p}+\frac{p'}{p}-
\frac{3}{x}\right)-\frac{6i\omega}{x}-
\frac{2\Lambda}{x^2}-\lambda\right\}R=0.
\end{equation}
Rescale $R$ as
\begin{equation}
R=x p^{-1} Y
\end{equation}
and insert it into equation \ref{c4e6}. After dividing everything by $x^2$, we
obtain
\begin{equation}\label{c4e8}
\left(-\partial_{x}-\frac{iq}{p}-\frac{4}{x}+\frac{p'}{p}\right)\left(\frac{3}
{x^2} Y-\frac{1}{x}Y'+\frac{iq}{xp}Y\right)-
\frac{6i\omega}{x^2p}Y-\frac{2\Lambda}{x^3 p}Y-\frac{\lambda}{xp} Y=0.
\end{equation}
Rescale $Y$ as
\begin{equation}\label{c4e7}
Y=e^{+i\omega r^{*}}Z,
\end{equation}
where
\begin{equation}\label{C430}
Z(x)=\sum_{n=0}^{\infty}a_{n}\frac{(x-x_{1})^{n}}{(-x_{1})^{n}}.
\end{equation}
By writing the radial solution in this way we achieve the result of factoring
out the behaviour of this function at the black hole event horizon $r=r_+\
(r_{*}=-\infty)$. Substitute $Y$ and its derivatives into \ref{c4e8}. We
finally get an equation in the function $Z$ that reads
\begin{equation}\label{C437}
Z''+\left(\frac{2iq}{p}-\frac{p'}{p}\right)Z'+\left(\frac{6i\omega}
{xp}-\frac{6}{x^2}-\frac{2\Lambda}{x^2p}-\frac{\lambda}{p}-\frac{3}
{x}\left(\frac{2iq}{p}-\frac{p'}{p}\right)\right)Z=0.
\end{equation}
The determination of the recurrence relation for the
radial series solution $Z$ can be found in appendix A of this paper.

\subsection{Angular equation}
The solution to the radial equation is strictly bound up with the determination
of the eigenvalues $\lambda(a,\alpha,m,\omega)$ of the angular equation, which
appear explicitly in the coefficients of the recurrence relation associated
with the series solution $Z$ of the radial differential equation. We could
attempt to solve for both $\omega$ and $\lambda$ simultaneously, but this
would be a time-consuming procedure. Instead, we shall attempt to determine
$\lambda$ before we find $\omega$. To this purpose, we first propose a second
order Taylor approximation of $\lambda$ for small $a\omega$. We then improve
this with the new numerical technique, already used  in the context of Kerr
black holes \cite{giammatteo}, based on a Pade approximating function for the
separation constant.

\subsubsection{Second order approximation for $\lambda$}
\label{soap}
We proceed in analogy with Sasaki \cite{sasaki}, who used a technique based on
perturbation theory to evaluate the separation constant for Kerr black holes
in asymptotically flat spacetime.
Our final result agrees with Sasaki's result when we assume the cosmological
constant to be equal zero.

We shall only consider $m=0$. Let us start by rewriting the angular equation in
the form
\begin{equation}
\left\{\left(\partial_{\mu}+\frac{K_{\mu}}{\Delta_{\mu}}-\frac{\Delta'_{\mu}}
{\Delta_{\mu}}\right)\Delta_{\mu}\left(\partial_{\mu}-\frac{K_{\mu}}
{\Delta_{\mu}}+\frac{\Delta'_{\mu}}{\Delta_{\mu}}\right)+6\omega
\mu-2\Lambda\mu^2+\lambda\right\}S=0.
\end{equation}
Since $m=0$, this implies for $K_{\mu}$ the expression
\begin{equation}
K_{\mu} =  a^2 \omega \sin^2 \theta.
\end{equation}
The equation can also be written as
\begin{equation}\label{c4e12}
HS = -\lambda S
\end{equation}
with
\begin{equation}
H=\Delta \partial_{\mu}^2+\Delta'\partial_{\mu}-K_\mu'+\Delta''-\frac{K_\mu^2}
{\Delta}+\frac{2k\Delta'}{\Delta}-\frac{\Delta'^{2}}
{\Delta}+6\omega\mu-2\Lambda\mu^2.
\end{equation}
First of all, it is straightforward to verify that when $\Lambda=0$ and $a=0$
we get from $H$ the expression we expect for static black holes in
asymptotically flat spacetime, as can be verified from \cite{chandra}.

We then introduce the new variable $u=\cos\theta$, such that $\mu =au$.
This gives 
\begin{equation}
H=\frac{\Delta}{a^2}\partial^2_{u}+\frac{\Delta'}
{a}\partial_{u}-K_\mu'+\Delta''-\frac{K_\mu^2}{\Delta}+\frac{2k\Delta'}
{\Delta}-\frac{\Delta'^2}{\Delta}+6\omega au-2\Lambda a^2 u^2.
\end{equation}

Next, we expand $H$ in powers of $a\omega$,
\begin{equation}\label{c4e9}
H=H_{0}+a\omega H_{1}+a^2\omega^2 H_{2}+O(a^3\omega^3),
\end{equation}
where,
\begin{eqnarray}
H_{0}&=&\frac{\partial^2}{\partial\theta^2}+\frac{\cos\theta}
{\sin\theta}\frac{\partial}{\partial\theta}+2-\frac{4}{\sin^2\theta}\\
H_{1}&=&-4\cos\theta\\
H_{2}&=&\frac{1}{3}\frac{\Lambda}
{\omega^2}\left(u^2-u^4\right)\partial^2_{u}+\frac{2}{3}\frac{\Lambda}
{\omega^2}\left(u-\frac{2}{3}u^3\right)\partial_{u}\nonumber\\
&+&\left\{-6\Lambda\frac{u^2}{\omega^2}+\frac{2}{3}\frac{\Lambda}
{\omega^2}-(1-u^2)+
\frac{(\frac{8}{3}\Lambda u^2-4\Lambda u^4)}{\omega^(1-u^2)}\right\}
\end{eqnarray}
We also expand $S$ and $\lambda$ in powers of $a\omega$ as
\begin{equation}\label{c4e10}
S = S_{0}+a\omega S_{1}+a^2\omega^2 S_{2}+O(a^3 \omega^3),
\end{equation}
\begin{equation}\label{c4e11}
\lambda = \lambda_{0}+a\omega
\lambda_{1}+a^2\omega^2\lambda_{2}+O(a^3\omega^3),
\end{equation}
with $\lambda_{0}=(l-1)(l+2)$.

Now, if we plug \ref{c4e9}, \ref{c4e10} and \ref{c4e11} into Eq. \ref{c4e12},
we have at the second order
\begin{eqnarray}
(H_{0}+a\omega H_{1}+a^2\omega^2H_{2})(S_{0}+a\omega S_{1}+a^2\omega^2
S_{2})=\nonumber\\-{(\lambda_{0}+a\omega\lambda_{1}+a^2 \omega^2 \lambda_{2})
(S_{0}+a\omega S_{1}+a^2\omega^2 S_{2})}.
\end{eqnarray}
Collecting the terms of the same order we obtain,
\begin{eqnarray}
(H_{0}&+&\lambda_{0})S^{l}_{0}  = 0\label{c4e14}\\
(H_{0}&+&\lambda_{0})S^{l}_{1}  = - (H_{1}+\lambda_{1})S^{l}_{0},
\hspace{3.0cm}\label{c4e15}\\
(H_{0}&+&\lambda_{0})S^{l}_{2}  =
-(H_{1}+\lambda_{1})S^{l}_{1}-(\lambda_{2}+H_{2})S^{l}_{0}.\label{c4e16}
\end{eqnarray}
The functions $S^{l}$ are the spheroidal harmonics of spin weight $s = -2$, for
$m=0$.

We will be interested in the explicit expressions of the functions $S^{l}_{0}$,
which  can be represented using Legendre Polynomials $P_{l}(\cos\theta)$, as
\begin{equation}
S^{l}_{0}(\theta) =  P_{l,\theta,\theta}-P_{l,\theta}\cot\theta.
\end{equation}
We should mention that at the moment the spherical harmonics are not
normalised. It is convenient to normalise at least those we will be using. The
normalised spherical harmonics are
\begin{eqnarray}
\hat S^{2}_{0} = \sqrt{\frac{5}{48}}S^{2}_{0}\\
\hat S^{3}_{0} =\sqrt{\frac{7}{240}}S^{3}_{0} \\
\hat S^{4}_{0} = \sqrt{\frac{1}{80}}S^{4}_{0}.
\end{eqnarray}
Since our purpose is to evaluate $\lambda$ up to the second order, let us start
with the calculation of $\lambda_{1}$ in Eq. \ref{c4e11}.

The first order correction to $\lambda$ is obtained by multiplying equation
\ref{c4e15} by $S^{l}_{0}$ from the left hand side and integrating it over
$\theta$. This gives
\begin{equation}
\lambda_{1}=-\int_{0}^{\pi}\hat S^{l}_{0}\;H_{1}\hat S^{l}_{0}\;\sin\theta
d\theta = 4\int_{0}^{\pi}(\hat S^{l}_{0})^2\;\cos\theta\sin\theta d\theta,
\end{equation}
but, if $m=0$, this integral is always zero for any value of $l$ as can,
indeed, be verified in \cite{sasaki}. 

Consequently,
\begin{equation}
\lambda_{1}=0, \;\;\;\;\;\forall l,\;\;\;m=0.
\end{equation}
To obtain $\hat S^{l}_{1}$, we set 
\begin{equation}
\hat S^{l}_{1}=\sum_{l'}c^{l'}_{l}\hat S^{l'}_{0} \hspace{2.0cm}m=0.
\end{equation}
We insert this expression into equation \ref{c4e15}, multiply it by $\hat
S^{l'}_{0}$ and integrate over $\theta$.
What we get is, 
\begin{equation}\label{c4e17}
c^{l'}_{l}(\lambda_{0}-\lambda'_{0})\int_{0}^{\pi}\left(\hat S^{l'}_{0}\right)
^2\sin\theta d\theta = -4\int_{0}^{\pi}d(\cos\theta)\hat S^{l'}_{0}\cos\theta
\hat S^{l}_{0}.
\end{equation}
The coefficients of the series on the left hand side, $c^{l'}_{l}$, can be
derived from \ref{c4e17} and written as
\begin{equation}
 c^{l'}_{l}=\frac{4}{(l'-1)(l'+2)-(l-1)(l+2)}
\int_{0}^{\pi}d(\cos\theta)\hat              
S^{l'}_{0}\cos\theta\hat S^{l}_{0}
\end{equation}
Hence $c^{l'}_{l}$ is generally non zero only for $l'= l\pm 1$.

The second order correction, $\lambda_{2}$, is obtained by multiplying equation
\ref{c4e16} by $\hat S^{l}_{0}$ form the left hand side and integrating it
over $\theta$. We find,
\begin{equation}
\lambda_{2}= -4\int_{0}^{\pi}d(\cos\theta)\hat S^{l}_{0}\cos\theta\hat
S^{l}_{1}+\int_{0}^{\pi}\hat S^{l}_{0}(H_{2}\hat S^{l}_{0}) d(\cos\theta).
\end{equation}
But 
\begin{equation}
\int_{0}^{\pi}d(\cos\theta)\hat S^{l}_{0}\cos\theta \hat
S^{l}_{1}=\frac14\sum_{l'}(c^{l'}_{l})^2[(l'-1)(l'+2)-(l-1)(l+2)]
\end{equation}
and the sum is non zero only for $l'=l\pm 1$.
It follows that,
\begin{eqnarray}\label{c4e18}
\lambda_{2}&=&-\left(c^{l+1}_{l}\right)^2
\left[l(l+3)-(l-1)(l+2)\right]\nonumber\\
&-&\left(c^{l-1}_{l}\right)^2\left[(l-2)(l+1)-(l-1)(l+2)\right]
+\int_{0}^{\pi}\hat S^{l}_{0}\;H_{2}\hat S^{l}_{0}\;d(\cos\theta).
\end{eqnarray}
If we, finally, evaluate the coefficients $c^{l'}_{l}$ and the integral on the
right hand side of \ref{c4e18}, when $l=2$, what we obtain for $\lambda_{2}$
is
\begin{equation}
\lambda_{2}= \frac{2}{7} \frac{\Lambda}{\omega^2}+\frac{10}{21}
\end{equation}
This gives us, for $l=2$, up to the second order in $a\omega$, an expression
for the separation constant $\lambda$ that reads
\begin{equation}
\lambda = \lambda_{0}+a^2 \omega^2\lambda_{2}= 4 +\frac{2}{7}\Lambda
a^2+\frac{10}{21}a^2 \omega^2.
\end{equation}
A similar calculation can be repeated for any other value of $l$.

For instance, we have performed the same calculation for $l=3$.
In that case, the Taylor expansion of $\lambda$ in the variable $a\omega$, up
to the second order is
\begin{equation}
\lambda=\lambda_{0}+a^2 \omega^2\lambda_{2}= 10 + \Lambda a^2+\frac{2}{3}a^2
\omega^2.
\end{equation}
In both cases, $l=2$ and $l=3$, we recover the flat space results for $\lambda$
as soon as we set $\Lambda=0$, as can be checked in \cite{sasaki}.

\subsubsection{Pade approximation for $\lambda$}
\label{pap}
Having showed how to obtain a second order approximation for the angular
eigenvalues, we would like to partially overcome the incompleteness of this
approach by a different scheme, which is based on the construction of a Pade
approximating function for $\lambda$. The technique which we describe below has
already been applied to find the quasi normal modes for Kerr black holes in
asymptotically flat space \cite{giammatteo}, 
and works well for all values of the parameter $a\omega$.

We will start from an analytical solution of the Teukolsky angular equation in
$AdS$ space-time. This equation is
\begin{equation}\label{c4e20}
\left(L_{-1}\Delta_{\mu}L^{\dagger}_{1}+6\omega\mu-2\Lambda\mu^2\right)S
=-\lambda S.
\end{equation}
If we replace the differential operators $L_{-1}$ and $L_1^{\dagger}$ with
their explicit expressions \ref{c4e19}, we can rewrite \ref{c4e20} as
\begin{equation}
S''+\frac{\Delta'_{\mu}}{\Delta_{\mu}}S'+\left( \frac{\Delta''_{\mu}}
{\Delta_{\mu}}-\frac{K_\mu'}{\Delta_{\mu}}-\frac{K_\mu^2}
{\Delta^2_{\mu}}+2K_\mu\frac{\Delta'_{\mu}}
{\Delta^2_{\mu}}-\frac{\Delta'^{2}_{\mu}}
{\Delta^{2}_{\mu}}+\frac{6\omega\mu}{\Delta_{\mu}}-\frac{2\Lambda\mu^2}
{\Delta_{\mu}}+\frac{\lambda}{\Delta_{\mu}}\right)S=0.
\end{equation}
This differential equation has a structure which is easily recognisable as
being very similar to the structure of the radial equation, with the variable
$\mu$ having assumed the role of $i$ times the radial variable $r$. It has
four regular singular points and they correspond to the four roots of
$\Delta_{\mu}$. They are $\mu=\pm a$ and $\mu=\pm \alpha$. 

As already done for the radial solution, it is possible to factorise out the
leading behaviour at $\mu=a$ by rescaling the angular function $S$ as
\begin{equation}
S=\Delta_{\mu}^{-1}e^{+i\omega\mu_{*}}\tilde Z,
\end{equation}
The variable $\mu_{*}$, which is analogous to the radial tortoise
coordinate $r_{*}$, is defined by
\begin{equation}
\frac{d}{d\mu_{*}}=\frac{\Delta_\mu}{(a^2-\mu^2)}\frac{d}{d\mu},
\end{equation}
while
\begin{equation}\label{c4e21}
\tilde Z= \sum_{n=0}^{\infty}a_{n}\frac{(\mu-a)^{n}}{(-2a)^{n}}.
\end{equation}
This allows us, through the same logic used to deal with the radial function
$R$, to get to this simpler form for the angular equation
\begin{equation}\label{c4e22}
\Delta_{\mu} \tilde Z''+\left(2K_\mu-\Delta'_{\mu}\right)\tilde
Z'+\left(6\omega\mu-2\Lambda\mu^2+\lambda\right)\tilde Z=0.
\end{equation}

Next, we expand all the coefficients of \ref{c4e22} as Taylor series around
$\mu = a$ and insert them into the equation itself beside the expression for
$\hat Z$ and its derivatives. We obtain a four term recurrence relation of the
form
\begin{equation}\label{c4e23}
\tilde\alpha_{n}a_{n+1}+\tilde\beta_{n}a_{n}+
\tilde\gamma_{n}a_{n-1}+\tilde\delta_{n}a_{n-2}=0.
\end{equation}

Finally, we  find the coefficients of \ref{c4e23} in the explicit form
\begin{eqnarray}
\tilde\alpha_{n}&=&(n^2-1)\left(1- a^2  \alpha^2\right)\\
\tilde\beta_{n}&=&-n \left(n-1\right)\left(1-5 a^2 
\alpha^2\right)+n\left(-4 a\omega+2-10 a^2 
\alpha^2\right)\nonumber\\
&& +6 a\omega+6 a^2\alpha^2+\lambda\\
\tilde\gamma_{n}&=&-8 a^2 
\alpha^2\left(n-1\right)\left(n-2\right)-2\left(n-1\right)\left(-2
a\omega-12 a^2 \alpha^2\right)\nonumber\\
&&-12 a\omega-24 a^2  \alpha^2\\
\tilde\delta_{n} & = & 4(n-4)(n-5) a^2 \alpha^2.
\end{eqnarray}
This allows us to determinate the coefficients $a_n$ of the series solution by
recursion.

Equation \ref{c4e22} has critical points at the north and south poles $\mu=\pm
a$. We require the solution which does not diverge at either pole. This
solution has $a_0=0$, $a_1=1$ and
\begin{equation}
\tilde Z=\sum_{n=0}^{\infty}a_{n}\frac{(\mu-a)^{n}}{(-2a)^{n}}\rightarrow 0 \
\;\;\hbox{as}\;\;\;\mu\rightarrow -a.
\end{equation}
Consequently, we have
\begin{equation}\label{c4e31}
\sum_{n=0}^{\infty} a_n=0.
\end{equation}
At this point, we wrote a {\em C++} computer program able to 
evaluate  values for $\lambda( a\alpha,a\omega)$, which are
identified with the roots of the left hand side of Eq. \ref{c4e31}, 
numerically. Within the program, the values
of $\lambda$ which satisfy Eq. (\ref{c4e31}) 
are found by the use of the Newton-Raphson method routine extended to the
complex plane \cite{press}. We have fit these numerical data for $\lambda$ to
Pade approximating functions for {\em real} $ a\omega$ at a fixed value of
$a\alpha$, given by
\begin{equation}
\lambda\sim \frac{a_0+a_1( a\omega)+a_2( a\omega)^2+a_3(
a\omega)^3}{1+b_1( a\omega)+b_2( a\omega)^2}.
\end{equation}
Finally, the Pade approximation and the numerical data are compared for {\em
imaginary} $ a\omega$. In the Pade formula, the numerator is one
degree higher than the denominator. The reason for this choice is that the
Teukolsky angular equation belongs to the family of differential equations for
angular prolate spheroidal wave functions, and the behaviour of the angular
eigenvalues for large $\omega$ is known \cite{abramovitz}.

\begin{figure}
\begin{center}
\leavevmode
\psfrag{Re(lambda)}{$\lambda_R$}
\psfrag{omegahat}{$\omega$}
\includegraphics{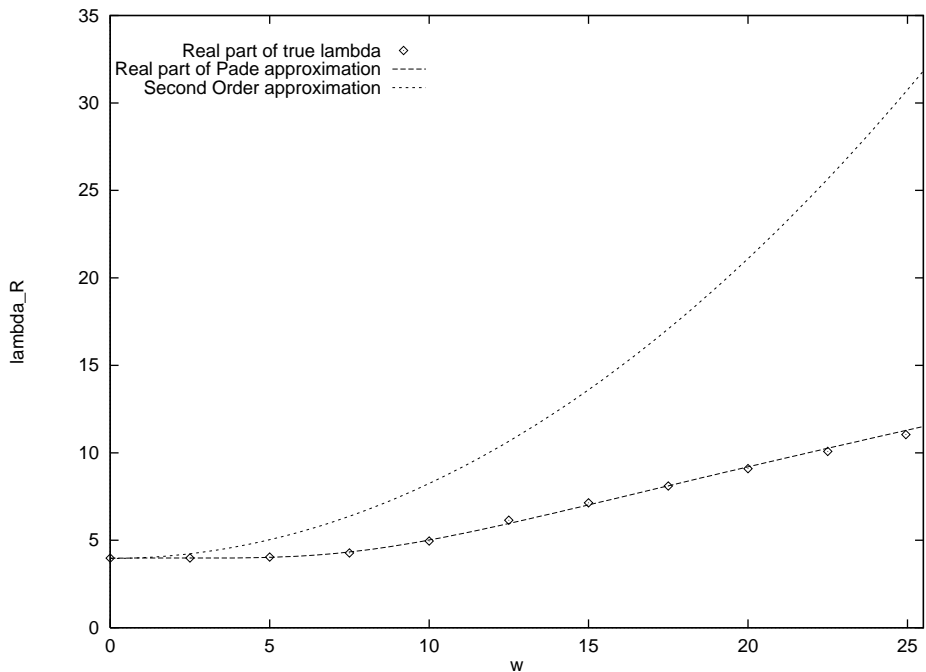}
\end{center}
\caption{The real part of the Pade approximation of the angular eigenvalue
$\lambda$ is shown to fit the data for lambda with imaginary $\omega$. Note
that the Pade approximation was constructed for real
$\omega$. The angular parameter  $ a \alpha= 0.24$, angular mode number $l=2$
and $m=0$. We also show the improvement we achieve with our Pade function
technique comparing our results to a second order approximation of $\lambda$.} 
\label{P1}
\end{figure}

\begin{figure}
\begin{center}
\leavevmode
\psfrag{omegahat}{$\omega$}
\psfrag{Im(lambda)}{$\lambda_I$}
\includegraphics{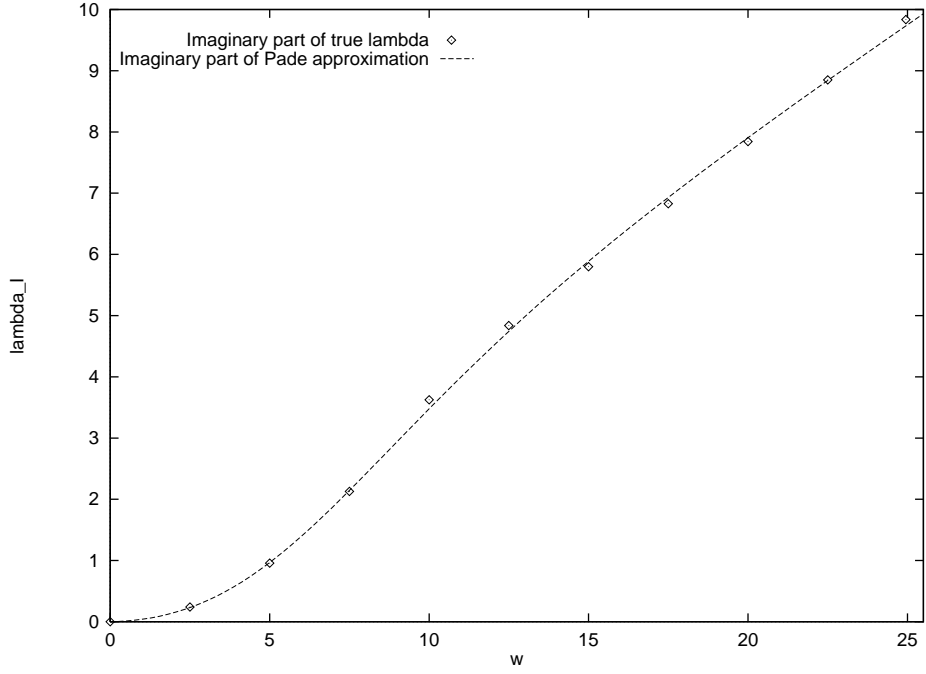}
\end{center}
\caption{The imaginary part of the Pade' approximation of the angular
eigenvalue $\lambda$ is shown, for $ a\alpha= 0.24$, $l=2$ and $m=0$.} 
\label{P2}
\end{figure}

In Fig. \ref{P1} and \ref{P2} we show how well the Pade function fits the data
for $\lambda$ coming from the computer program. We also show the improvement
compared to the second order approximation and we plot results for the real
and imaginary parts of $\lambda$.
The fit parameters of the rational function representing $\lambda$ are shown in
Table \ref{T1} for values of the angular parameter $a\alpha$ ranging between
$0$ and $0.4$, for the $l=2$ angular mode number.
\begin{table}[ht!]
\begin{center}
\begin{tabular}{c|cccccc}
\hline
\hline
 Parameter $\alpha a$&
$a_{0}$&$a_{1}$&$a_{2}$&$a_{3}$&$b_{1}$&$b_{2}$\\\hline     
0.00&4.0&0.0&0.0&0.0&0.0&0.0\\\hline                        
0.08&4.0&0.0560&0.6068&0.0090&0.0139&0.0329\\ \hline
0.16&4.0&0.5350&0.7121&0.1039&0.1307&0.0641\\ \hline
0.24&4.0&0.9169&0.8387&0.2233&0.2177&0.1219\\ \hline
0.32&4.0&1.1610&0.9411&0.3368&0.2658&0.1770\\ \hline
0.40&4.0&1.4410&1.1728&0.3670&0.3936&0.1916\\ \hline
\hline
\end{tabular}
\caption{This table shows the parameters we obtain by fitting the Pade
approximation for $\lambda$, to the numerical data coming out from the
computer program for the calculation of the angular eigenvalues of the
Teukolsky angular equation. These results have been obtained by fixing the
angular mode number $l=2$ and $m=0$.}
\label{T1}
\end{center}
\end{table} 

\section{Transformation theory and the Regge-Wheeler equation}
\label{rbcond}
We have to impose some boundary
conditions on the radial equation before we can obtain an equation for the
frequencies of the quasinormal modes.  We shall do this by extending the
transformation theory of perturbations of black holes to AdS space, allowing us
to construct the Regge-Wheeler-Zerilli equation for the underlying the metric
perturbations.

The transformation theory for Kerr black holes has been described in detail by
Chandrasekhar \cite{chandra}. Following the Kerr case, we reparameterise the
Weyl function $\Psi_0$ by
\begin{equation}
\Psi_0=e^{+i\omega t}\frac{\varpi^3}{\Delta}Y(r)S_l(\theta).
\end{equation}
where
\begin{equation}
\varpi^2=r^2+\widetilde a^2=r^2+a^2+{am\chi^2\over\omega}
\end{equation}
We also adopt a new $r_*$ coordinate,
\begin{equation}
dr_*={\varpi^2\over \Delta_r}dr
\end{equation}
The Teukolsky equation for $Y$ becomes
\begin{equation}
\left\{\left(\partial_*^2+\omega^2\right)
+P\left(\partial_*-i\omega\right)-Q\right\}Y=0\label{teq}
\end{equation}
where $Q=\Delta_r u/\varpi^8$, and $u$ is a polynomial
\begin{equation}
u=(\lambda-6\alpha^2r^2)\varpi^4+
3(r\Delta_r'-\Delta_r)\varpi^2-3r^2\Delta_r
\end{equation}

Part of the aim of transformation theory is to construct a function $\psi$
which satisfies the Regge-Wheeler-Zerilli equation
\begin{equation}
\left(\partial_*^2+\omega^2\right)\psi=V\psi\label{zeq}
\end{equation}
with a potential function $V(r)$. The function $\psi$ represents the metric
perturbations of the black hole.

The two radial functions $Y$ and $\psi$ are related by an expression of the
form
\begin{equation}
{\varpi^8\over\Delta_r^2}R\,Y-
{\varpi^8\over\Delta_r^2}T\left(\partial_*-i\omega\right)Y,
=K\psi\label{c4e32}
\end{equation}
where $R$ and $T$ are functions of $r$ and $K$ is a constant. (In this section,
we follow Chandrasekhar's notation and $T$ does not denote the temperature.)
The consistency of the two equations (\ref{teq}) and (\ref{zeq}) allows one to
derive expressions for $R$, $T$ and $K$,
\begin{eqnarray}
R&=&{\Delta_r\over \varpi^8}(u+\beta\Delta_r)\\
T&=&{v-\kappa_2\Delta_r\varpi^2
\over u-\beta\Delta_r}+2i\omega\\
K&=&\kappa-4\omega^2\beta+2i\omega\kappa_2
\end{eqnarray}
where
\begin{equation}
v=u'\Delta_r-u\Delta_r'
\end{equation}
Finding the constants $\kappa$, $\kappa_2$ and $\beta$ requires a substantial
amount of algebra\footnote{A MAPLE worksheet is available from the authors}.
Some help can be obtained by using the Starobinski constant
$|K|^2$ for Kerr-AdS given in \cite{chambers}. We obtain,
\begin{eqnarray}
\kappa&=&\lambda(\lambda+2+2\alpha^2a^2)-12\alpha^2a^2\\
\kappa_2&=&\pm(36M^2+2\beta\kappa-12\lambda\widetilde a^2-10\lambda^2\widetilde
a^2
+24\lambda a^2\nonumber\\
&&+24\lambda\widetilde a^4\alpha^2-12\lambda\widetilde a^2a^2\alpha^2
-72\widetilde a^2 a^2\alpha^2)^{1/2}\\
\beta&=&\pm3\widetilde a^2
\end{eqnarray}
There are four consistent solutions corresponding to the choice of signs in the
above expressions.

The final ingredient in the transformation theory is the potential,
\begin{equation}
V=\Delta_r\left\{
{(\kappa_2\varpi^2\Delta_r-v)
(\kappa_2\varpi^2u-\beta v)\over
\varpi^4(u+\beta\Delta_r)(u-\beta\Delta_r)^2}
+{\kappa\over u+\beta\Delta_r}
-\beta{\Delta_r\over \varpi^8}\right\}
\end{equation}
There are four different potentials depending on the choice of signs for
$\kappa_2$ and $\beta$. Some general features of the potential are that it
vanishes as $r\to r_+$ and approaches a non-zero constant as $r\to \infty$
($r_*\to 0$). The basic situation can be interpreted as  scattering from
$r_*\to-\infty$  to a boundary at $r_*=0$. 

The potential has poles at $r=|\tilde a|$ for real $\omega<\omega_s$, where 
\begin{equation}
\omega_s=-{am\chi^2\over r_+^2+a^2}.
\end{equation}
(The potential can also have poles where $u\pm\beta\Delta_r=0$.)
Note that the $r_*$ variable is not a single valued function of $r$ when
$\omega<\omega_s$. This frequency range corresponds to the super-radiant
regime of Kerr. For kerr-AdS, the existence of super-radiant frequencies can
be prevented by the boundary conditions at $r_*=0$.

In the axisymmetric case, $m=0$, the potential is always continuous and
bounded. As with Kerr, two of the potentials can be complex when $\kappa_2$ is
imaginary.

Suppose that we are interested in the evaluation of some correlation function
of a polar quantity in the conformal field theory which resides on the
boundary of our $AdS$ space. The $AdS/CFT$ correspondence conjectures that
this polar quantity can be associated to the polar metric perturbation of an
$AdS$ black hole living in the bulk. Accordingly we must set the axial black
hole perturbation to vanish on the boundary. Since the axial perturbation
function is the Regge-Wheeler function $\psi$ with $\kappa_2<0$, this also
means that we are led to assume that this $\psi$ goes to zero as $r$ goes to
infinity.

On the other hand, we have series solutions for $Y(r)$. Let us consider
the boundary conditions on this function. First of all, we have from Eq.
\ref{c4e32}
\begin{equation}
RY-T\left(\partial_*-i\omega\right)Y\rightarrow 0.
\end{equation}
We find, for $\beta>0$, 
$T\longrightarrow T_\infty=
2i\omega-(\kappa_2+6M)/(\lambda-6\alpha^2\tilde a^2)$
and  $R\longrightarrow \alpha^2\lambda$ as $r\rightarrow\infty$. 
The differential operator $\partial_{r_{*}}$ as $r\rightarrow\infty$ yields
$\partial_{r_{*}}\longrightarrow-\alpha^2\partial_{x}$ (recall $x=r^{-1}$).
If we rescale $Y$ as 
\begin{equation}
Y= e^{+i\omega r_{*}}Z ,
\end{equation} 
with $Z$ representing once again the series solution, then we find
\begin{equation}
\left(\frac{\lambda}{T_\infty}\right)Z+Z'=0,
\end{equation}
which represents the boundary condition for the axial problem we are
considering. This equation tells us that assuming Dirichlet boundary condition
on the Regge-Wheeler function $\psi$, we are left with Robin conditions on the
solution of our radial Teukolsky equation. Moreover, since
\begin{equation}
Z\longrightarrow \sum_{n=0}^{\infty}a_{n},\hbox{ and } 
Z'\longrightarrow -r_+ \sum_{n=0}^{\infty} \left(n+1\right)a_{n+1},
\end{equation}
as $x\rightarrow 0$ ($r \rightarrow \infty$), the boundary conditions can be
explicitly written in the form
\begin{equation}
\frac{\lambda}
{r_+T_\infty}\sum_{n=0}^{\infty}a_{n}-
\sum_{n=0}^{\infty}\left(n+1\right)a_{n+1}=0,\label{bcz}
\label{bcond}
\end{equation}
In practice, $T_\infty\approx 2i\omega$, and the numerical solutions described
in the next section have used $T_\infty=2i\omega$, in which case the boundary
conditions reduce to those used in \cite{moss}.

The transformation theory which we have obtained is sufficiently powerful to
give us the frequencies of some special quasi normal modes. The Teukolski
function $Y$ gives rise to four independent radial functions, from which we
select $\psi^\pm$ with oposite signs for $\kappa_2$.  Working back from eq.
(\ref{c4e32}), we can relate the radial functions by
\begin{equation}
K^\pm\psi^\pm=C^\pm\psi^\mp+D^\pm\partial_*\psi^\mp
\end{equation}
for some functions $C^\pm$ and $D^\pm$. Now we take $\omega$ to be a root of
$K^+(\omega)=0$, and solve for $\psi^-$ to obtain a quasi normal mode. The
frequencies are
\begin{equation}
\omega={i\over 4\beta}\left(\kappa_2\pm(\kappa_2^2-4\beta\kappa)^{1/2}\right)
\label{spec}
\end{equation}
Note that this is an implicit equation in the rotating case,  since the
constants depend on $\omega$. In the non-rotating case, the frequencies reduce
to $\omega=i\kappa/12\pi M$, which was found in \cite{moss}. In the rotating
case, the frequencies can be pure imaginary or complex. These values provide a
useful check on the accuracy of the numerical results presented in the next
section.

\section{numerical results}
\label{numres}
In this section we will present some numerical results obtained by using the
procedures outlined above for the solution of both the radial and the angular
Teukolsky equations. The radial boundary conditions used and the motivations
for choosing them are those specified in section \ref{rbcond}. We replace the
angular eigenvalue $\lambda$ by the Pade approximation and then the
quasi-normal frequencies $\omega$ are roots of eq. (\ref{bcz}).

\begin{figure}[ht!]
\begin{center}
\leavevmode
\includegraphics{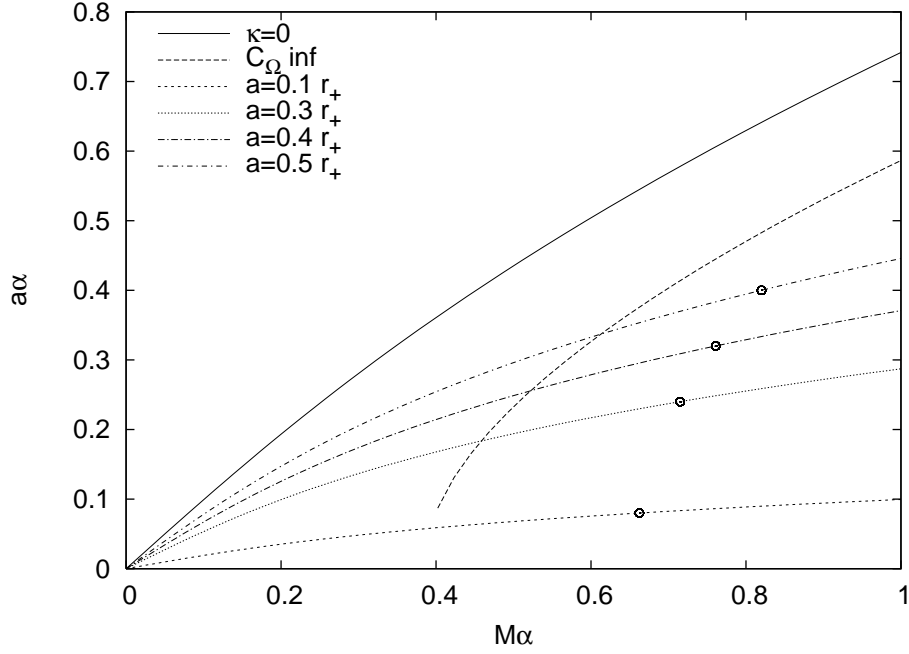}
\end{center}
\caption{The parameter ranges used for the intermediate size of black hole lie
along the lines shown in the figure. The small circles show the parameters
used in Fig. \ref{freq1}} 
\label{pars}
\end{figure}

Small Kerr Anti de-Sitter black holes, with the
radial coordinate of the event horizon $r_+$ much smaller than the AdS radius
$\alpha^{-1}$  $(r_+\ll\alpha^{-1})$, represent an unstable thermodynamic
phase. Hence, we concentrate on larger black hole radii.
In order to avoid metric singularities, the black hole rotation parameter is
constrained to be less than the AdS radius, $a < \alpha^{-1}$.
The black hole parameters which we have used are displayed in Fig. \ref{pars}.

\begin{figure}[ht!]
\begin{center}
\leavevmode
\psfrag{wI}{$\hat \omega_I$}
\psfrag{wR}{$\hat \omega_R$}
\includegraphics{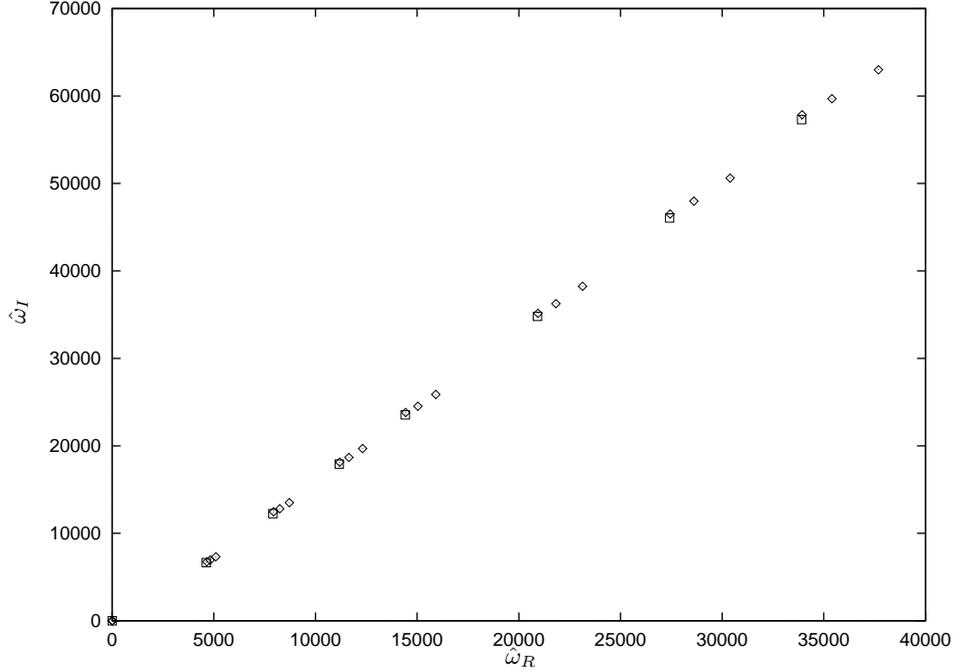}
\end{center}
\caption{The frequencies of the first few quasinormal modes for a Kerr black
hole in Anti-de Sitter metric with $r_+ = 50\;\alpha^{-1}$. Results are
plotted for the angular parameter $\alpha a = 0.0, 0.1, 0.2, 0.3$. The
squares represent the results for Schwarzschild $AdS$ black holes. The
frequencies are rescaled by $r_+$, so that $\hat\omega=\omega r_+$.} 
\label{freqlbh}
\end{figure}

\begin{figure}[ht!]
\begin{center}
\leavevmode
\psfrag{wI}{$\hat \omega_I$}
\psfrag{wR}{$\hat \omega_R$}
\includegraphics{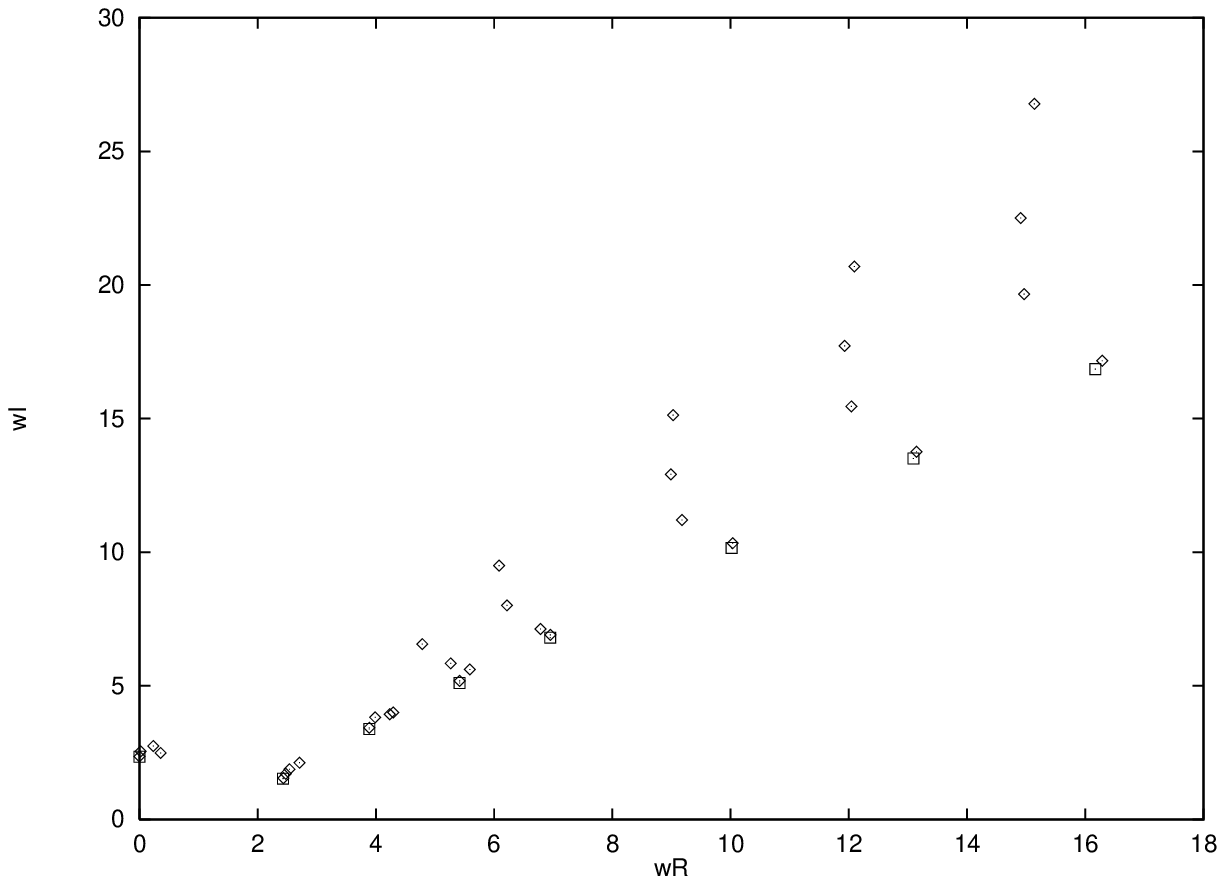}
\end{center}
\caption{The frequencies of the first few quasinormal modes for a Kerr black
hole in Anti-de Sitter metric with $r_+ = 0.8\;\alpha^{-1}$. Results are
plotted for the angular parameter $ \alpha a = 0.0, 0.08, 0.24, 0.32, 0.40$.
The squares represent the results for Schwarzschild $AdS$ black holes. The
frequencies are rescaled by $r_+$, so that $\hat\omega=\omega r_+$.} 
\label{freq1}
\end{figure}

The quasinormal frequencies are decomposed into real and imaginary parts,
$\omega=\omega_{R}+i\omega_{I}$. Having chosen the metric perturbation with
time dependence $e^{+i\omega t}$, we have that the pure imaginary component
$\omega_{I}$ is positive for all quasinormal frequencies.  Fig. \ref{freqlbh},
shows the evolution of the quasinormal frequencies of large black holes (LBH),
with respect to the angular parameter $ a$, for $r_+=50\;\alpha^{-1}$.
Fig. \ref{freq1} shows the evolution of the quasinormal frequencies of
intermediate size black holes (IBH), with respect to the angular parameter
$a$, for $r_+=0.8\ \alpha^{-1}$. 

To obtain both pictures, we have used the Pade approximation for $\lambda$ and
therefore we have been free to consider frequencies as large as we liked. The
squares represent the results for static $AdS$ black holes. Unfortunately, the
numerical procedure described in section \ref{pap} becomes unreliable for
values of $a$ higher than $0.3\alpha^{-1}$ (for LBH) and $0.4\alpha^{-1}$ (for
IBH), precluding us from extending the present results to very rapidly
rotating holes. 

The quasinormal frequencies for rotating black holes show a behaviour which is
very similar to the one found in the case of Schwarzschild black holes
\cite{konoplya}. They have a real part which displays an increase along with
the imaginary part. The spacing between consecutive modes tends to become
regular as the overtone number $n$ gets larger, and it seems to depend on the
angular parameter $a$. Since in $AdS$ space the first ten frequencies are
usually enough to ascertain their asymptotic behaviour, we have computed the
quasinormal frequencies for $\alpha a=0.1, 0.2, 0.3$ in the case of LBH, and
$\alpha a = 0.08, 0.24, 0.40$ in the case of IBH, up to the tenth mode. What
we have obtained for the spacing in the LBH regime is,
\begin{eqnarray}
\omega_{n+1}-\omega_{n}&\sim 
&(64.3+112\;i)\alpha\hspace{1.0cm}\alpha a
=0.1,\\
\omega_{n+1}-\omega_{n}&\sim 
&(65.2+113\;i)\alpha\hspace{1.0cm}\alpha a
=0.2,\\
\omega_{n+1}-\omega_{n}&\sim 
&(66.5+113\;i)\alpha\hspace{1.0cm}\alpha a
=0.3.
\end{eqnarray}
while in the IBH regime we have,
\begin{eqnarray}
\omega_{n+1}-\omega_{n}&\sim &
(1.95+2.11\;i)\alpha\hspace{1.0cm}\alpha a
=0.08
\\
\omega_{n+1}-\omega_{n}&\sim &
(1.72+2.46\;i)\alpha\hspace{1.0cm}\alpha a
=0.24
\\
\omega_{n+1}-\omega_{n}&\sim &
(1.57+2.85\;i)\alpha\hspace{1.0cm}\alpha a
=0.40.
\end{eqnarray}
For the smaller hole, there appears to be evidence that the real part of the
spacing decreases with $a$ whilst the imaginary part of the spacing increases
with $a$. From Figs. \ref{pars} and \ref{C4F4} we see that the temperature is
increasing with $a$ in this region of parameter space.

\begin{figure}[!ht]
\begin{center}
\leavevmode
\psfrag{k/alpha}{$\kappa\alpha^{-1}$}
\psfrag{Re(w)/alpha}{$\omega_R \alpha^{-1}$}
\scalebox{0.7}{\includegraphics{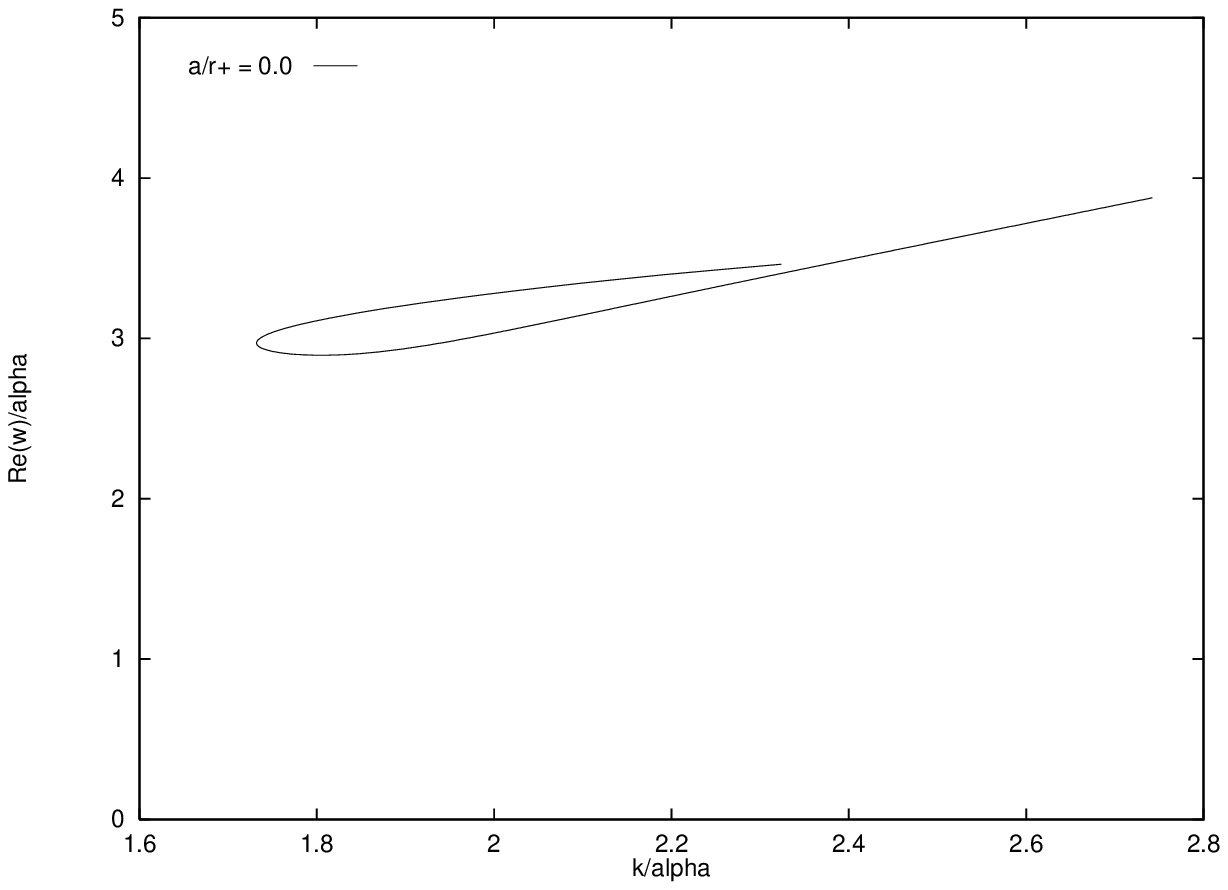}}
\scalebox{0.7}{\includegraphics{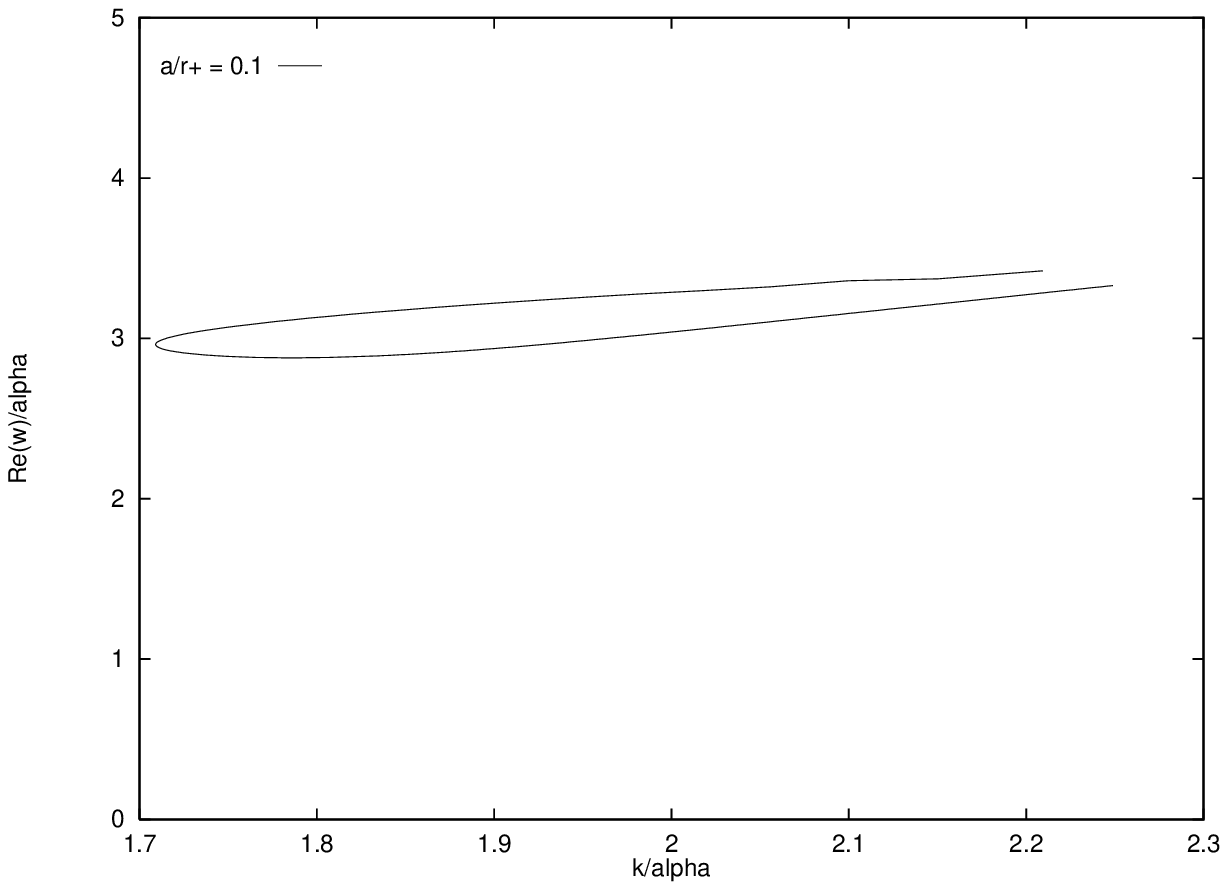}}
\scalebox{0.7}{\includegraphics{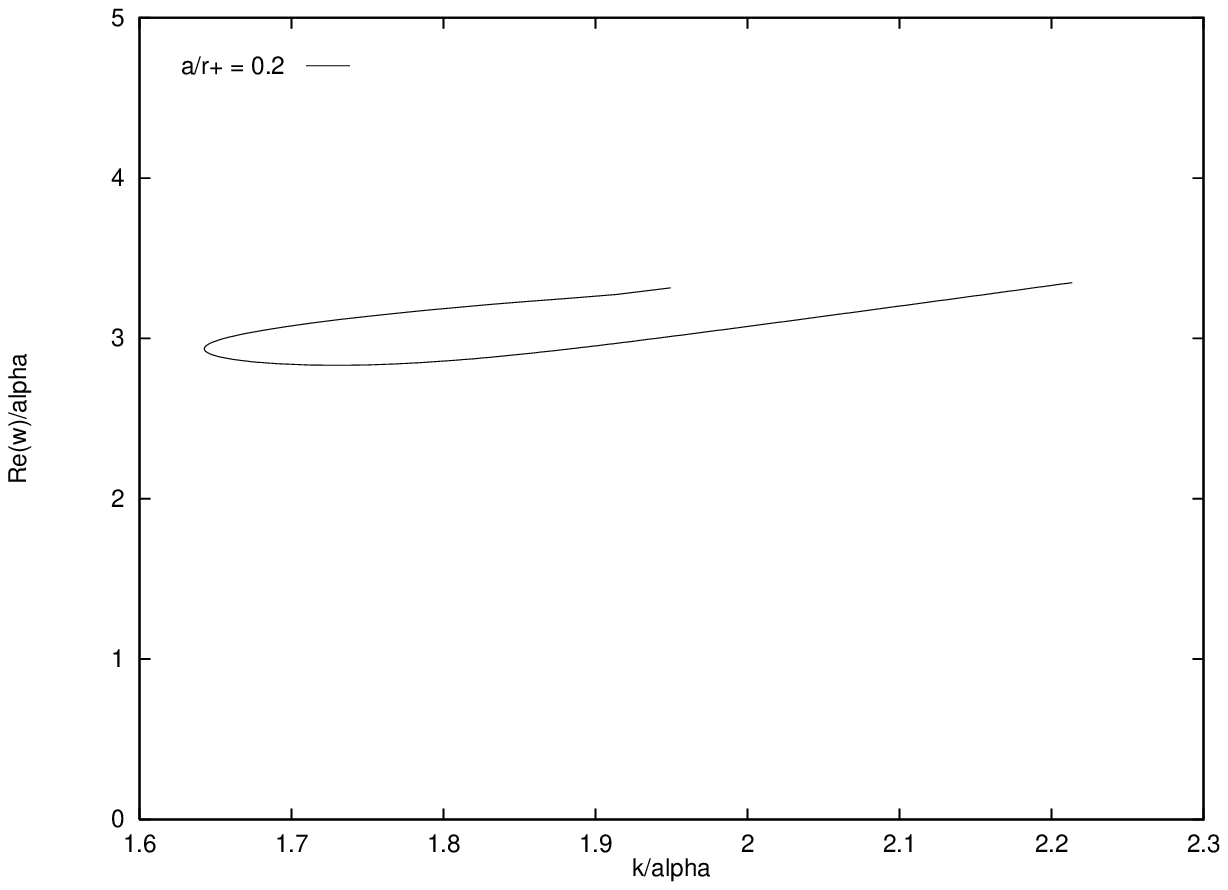}}
\scalebox{0.7}{\includegraphics{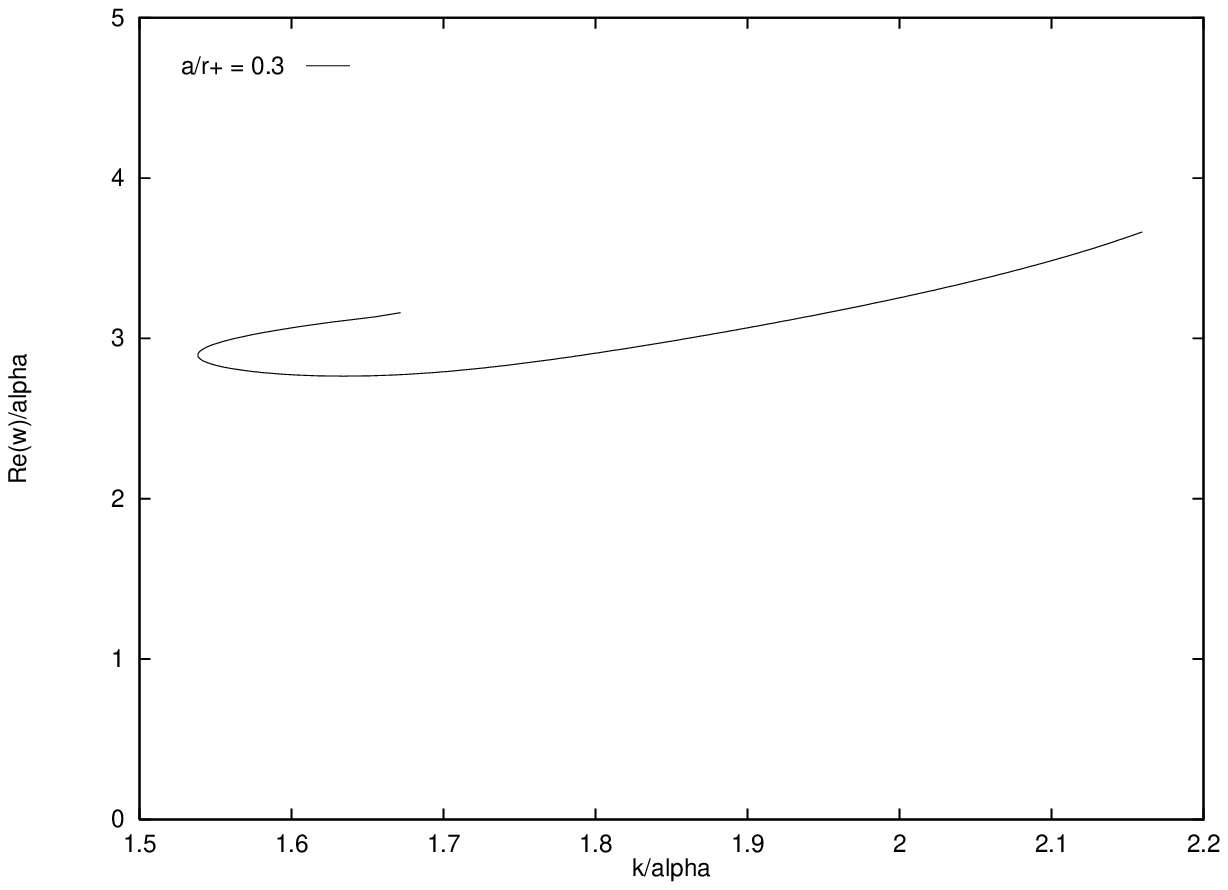}}
\end{center}
\caption{The dependence of the real part of the first non-special quasinormal
mode frequency on the metric parameters is shown for $l=2$ and $a/r_+=0.0,
0.1, 0.2, 0.3$, where $r_+$ is the radius of the event horizon, $\kappa$
represents the surface gravity of the black hole and
$\alpha^{-1}$ is the Anti-de Sitter radius.} 
\label{Par1}
\end{figure}

\begin{figure}[!ht]
\begin{center}
\leavevmode
\psfrag{k/alpha}{$\kappa\alpha^{-1}$}
\psfrag{Im(w)/alpha}{$\omega_I \alpha^{-1}$}
\includegraphics{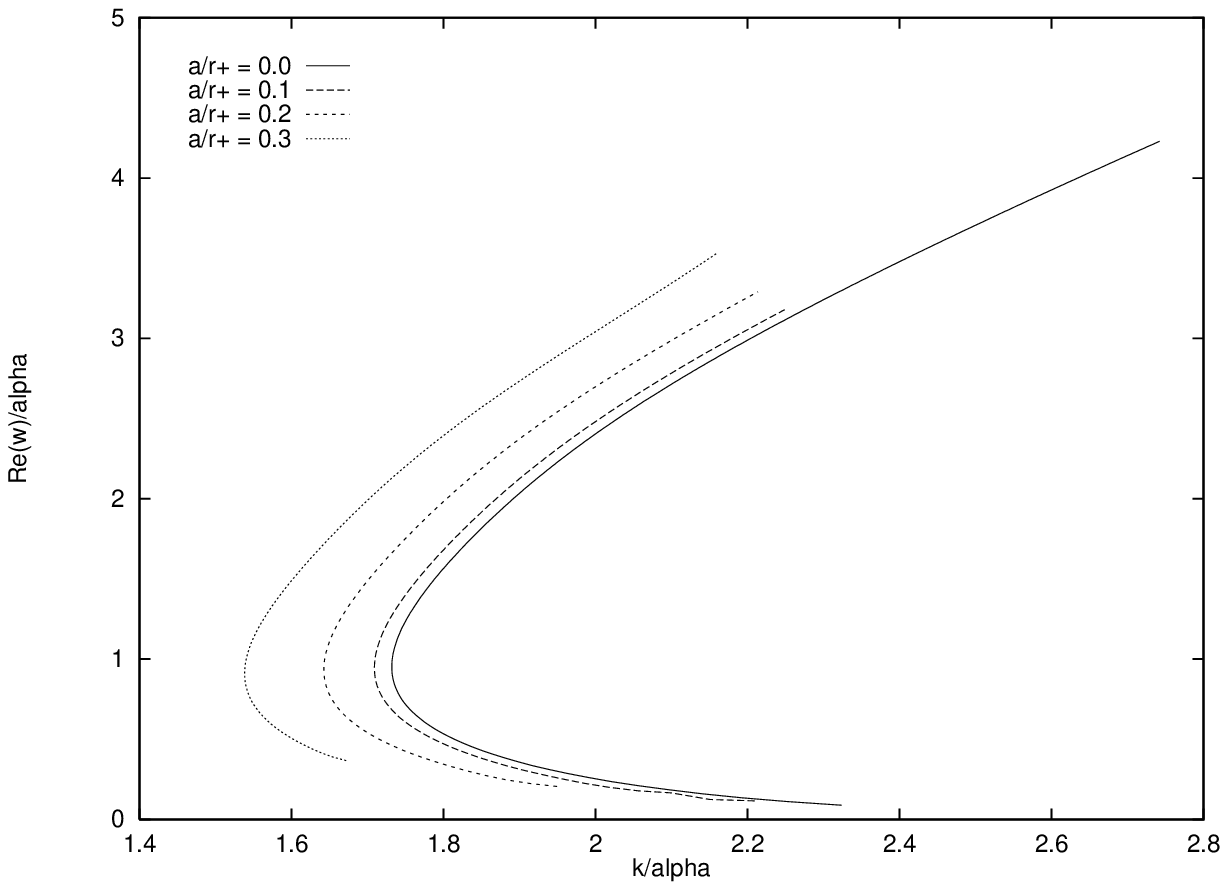}
\end{center}
\caption{The dependence of the imaginary part of the first non-special
quasinormal
mode frequency on the metric parameters is shown for $l=2$ and $ a/r_+=0.0,
0.1, 0.2, 0.3$, where $r_+$ is the radius of the event horizon, $\kappa$
represents the surface gravity of the black hole and
$\alpha^{-1}$ is the Anti-de Sitter radius. The upper branch corresponds to
larger values of $r_+$ than the lower branch. } 
\label{Par2}
\end{figure}

Both the real and imaginary parts of the frequencies become proportional to the
surface gravity $\kappa$ for large $\kappa$, which agrees with the behaviour
for static black hole quasinormal modes \cite{horowitz, moss}. This is
displayed in figures \ref{Par1} and \ref{Par2} in which we show the dependence
of the real and imaginary parts of the first $(n=1)$ quasinormal frequency on
the metric parameters for some values of $\hat a$. Note that these
pictures have been obtained by exploiting the second order approximation for
$\lambda$,
and that the results for $\hat a =0$ are those obtained by Moss and Norman in
\cite{moss}.

The surface gravity reaches a minimum close to the critical line in Fig.
\ref{pars} on which $C_\Omega$ diverges. There is some numerical evidence in
figures
\ref{Par1} and \ref{Par2} that the quasi normal frequencies are constant along
the critical line, with $\omega\approx(3+i)\alpha$.

\section{Discussion}
There have been investigations on $AdS$ black holes particularly focused on
their thermodynamic properties since the beginning of the eighties
\cite{page}.
Recently, Anti-de Sitter space has appeared in the surprising new context of
the Maldacena conjecture relating the physics of $AdS$ in $n+1$ dimensions to
a strongly coupled conformal (gauge) field theory $(CFT)$ in one dimension
fewer. This interest has led to the study of static, charged and rotating black
holes in $AdS$ space in five and even seven dimensions, where the $AdS/CFT$
correspondence can be still formulated. References on these investigations can
be found, for instance, in \cite{banados1, banados2, hunter, reall,
birmingham2, johnson, awad}.

Large Kerr $AdS$ black holes are stable against Hawking evaporation. This
property is not shared by asymptotically flat Kerr black holes. Furthermore,
Kerr-$AdS$ black holes can have rotating Killing vectors which are time-like
everywhere outside the horizon. Following the orbits of these rotating Killing
vectors, thermal radiation can rotate around the black hole with the same
angular velocity of the hole, implying that the phenomenon known as
super-radiance is not allowed, and the black hole is in thermal equilibrium
with the thermal gas around it \cite{hunter, reall}.

Although research has been carried on for a few years in the field of
quasinormal modes of black holes in $AdS$ space and results have been obtained
for Schwarzschild and Reissner-Nordstrom black holes, it is in this work that
for the first time that frequencies of Kerr $AdS$ black holes are computed.
The possibility is that, through the $AdS/CFT$ correspondence, such
frequencies can be exploited in the calculation of the relaxation times of the
perturbation of the corresponding thermal state in the conformal field theory
on the boundary.  

It would be desirable to analyse the Kerr $AdS$ black holes
in deeper detail and see how the frequencies relate to the thermodynamical
properties of the black holes and whether there is a direct interpretation in
terms of the $CFT$. In particular, it would be interesting to look for special
behaviour along the phase transition line and to examine the non-axisymmetric
case to find out what happens to the super radiant modes.

It is also worth mentioning that, although black hole solutions in higher
dimensions than $AdS_4$ have been found \cite{johnson, awad}, the calculation
of their quasinormal frequencies has never been done. Therefore an extension
of our work to $AdS_{5}$ and $AdS_{7}$, where the $AdS/CFT$ correspondence can
be still formulated, seems possible.

Finally, it would be perhaps of some interest to further check if the ideas on
black hole area quantisation and the relationship to highly damped quasinormal
frequencies, can be translated into non-asymptotically flat spacetimes, as
Anti-de Sitter space.

\appendix

\section{Recurrence relation for the series solution $Z$}
In this appendix we describe in details how to derive the recurrence relation
of the series solution $Z$ for the radial Teukolsky equation associated to
spin-$2$ perturbations of Kerr Anti-de Sitter black holes.
  
Starting from equation \ref{C430} for $Z$, we can write its derivatives with
respect to $x$ as
\begin{eqnarray}
Z' &=&\sum_{n=1}^{\infty}a_{n}(-x_{1})^{-n}n(x-x_{1})^{n-1}\nonumber\\
Z''&=&\sum_{n=2}^{\infty}a_{n}(-x_{1})^{-n}n(n-1)(x-x_{1})^{n-2}. \nonumber
\end{eqnarray}
We re-write Eq. \ref{C437} as
\begin{equation}\label{A4}
pZ''+\left(2iq-p'\right)Z'+\left(6ia^2 x  \omega -6Mx+6a^2 x^2
-\lambda\right)Z=0.
\end{equation}
Next, consider expressions \ref{C425} and \ref{C426} for $q(x)$ and $p(x)$ and
call $C(x)$ and $D(x)$ the coefficients of $Z'$ and $Z$ in Eq. \ref{A4},
\begin{equation}
p(x)Z''+C(x)Z'+D(x)Z=0,
\end{equation}
and let us develop all these coefficients as Taylor series around
$x=x_{1}=1/r_+$, which is the inverse of the radial coordinate of
the event horizon of the black hole. We have,
\begin{equation}
p(x)= \alpha^2+(1+a^{2}\alpha^2)x^2-2Mx^3+a^{2}x^4,
\end{equation}
\begin{equation}
C(x)=-2i\omega(1+a^2x^2)-p'(x)
\end{equation}
\begin{equation}
D(x)=6ia^2x \omega-6Mx+6a^2 x^2 -\lambda 
\end{equation}
Note that $p(x_{1})=0$. In fact $p(x)=\Delta_r x^4$ and $x_{1}$ is, by
definition, the largest root of $\Delta_{r}(x)$.

When, in Eq. \ref{A4}, we replace its coefficients with their Taylor expansions
and insert into the same equation the series expressions for $Z$ and its first
and second derivatives, we obtain the recurrence relation,
\begin{equation}
\alpha_{n}a_{n+1}+\beta_{n}a_{n}+\gamma_{n}a_{n-1}+\delta_{n}a_{n-2}=0,
\end{equation}
where the coefficients are given by
\begin{eqnarray}
\alpha_{n}&=&(-x_{1})^{-(n+1)}\left[p'(x_{1})n(n+1)+C(x_{1})
(n+1)\right]\nonumber\\
\beta_{n}&=&(-x_{1})^{-n}\left[\frac{1}{2}p''(x_{1})n(n-1)+C'(x_{1})
(n-1)+D(x_{1})\right]\nonumber\\
\gamma_{n}&=&(-x_{1})^{-(n-1)}\left[\frac{1}{6}p'''(x_{1})(n-1)(n-2)+\frac{1}
{2}C''(x_{1})(n-1)+D'(x_{1})\right] \nonumber\\
\delta_{n}&=&(-x_{1})^{-(n-2)}\left[a^2(n-2)(n-3)-4a^2
(n-2)+6a^2\right]
\end{eqnarray}
and we impose the initial conditions, $a_{-2}= a_{-1} = 0$ and $a_{0}=1$. Note
that
\begin{equation}
p'(x_1)=-2(1+a^2x_{1}^2)\kappa
\end{equation}
where $\kappa$ is the surface gravity.

We re-scale all the parameters with respect to $x=x_{1}$, so that $\hat a=ax_1$
etc. The re-scaled mass
\begin{equation}
\hat M=\frac12(1+\hat a^2)(1+\hat\alpha^2)
\end{equation}
and the re-scaled surface gravity 
\begin{equation}
\hat\kappa={1+3\hat\alpha^2-\hat a^2+\hat\alpha^2\hat a^2
\over 2(1+\hat a^2)}.
\end{equation}
Substituting everything into the equations for the coefficients of our
recurrence relation, we finally obtain
\begin{eqnarray}
\alpha_{n}&=&2(n^2-1)\hat\kappa(1+\hat a^2)+
(n+1)2i\hat \omega\,(1+\hat a^2),\\
\beta_{n}&=&n(n-3)(6\hat a^2-6\hat M+1+\hat a^2\hat
\alpha^2)-2(n-3)\,2i\hat\omega\,\hat a^2+
(6\hat a^2-6\hat M-\lambda),\\
\gamma_{n}&=&(n-2)(n-4)(2\hat M-4\hat a^2)
+(n-4)2i\hat\omega\,\hat a^2\nonumber,\\
\delta_{n}&=&(n-4)(n-5)\hat a^2.
\end{eqnarray}

Notice is that, since we are
considering Anti-de Sitter space with a negative
cosmological constant $\alpha^2>0$, 
\begin{equation}
\hat \kappa>\frac{(1-\hat a^2)}{2(1+\hat a^2)}.
\end{equation}
This is the same limit mentioned in \cite{moss} for Schwarzschild Anti-de
Sitter black holes.


\begin{thebibliography}{99}
\bibitem{hooft}
G. 't Hooft, gr-qc/9310026 (1993)
\bibitem{susskind}
L. Susskind, J. Math. Phys. {\bf 36}, 6377 (1995)
\bibitem{maldacena}
J. Maldacena, Adv. Theor. Math. Phys. {\bf 2}, 231 (1998)
\bibitem{gubser}
S.S. Gubser, I. R. Klebanov, A. M. Polyakov, Phys. Lett. B {\bf 428}, 105
(1998)
\bibitem{witten}
E. Witten, Adv. Theor. Math. Phys. {\bf 2}, 253 (1998)
\bibitem{aharony}
O. Aharony, S.S. Gubser, J. Maldacena, H. Ooguri, Y. Oz, Phys. Rept. {\bf 323},
183 (2000)
\bibitem{petersen}
J. L. Petersen, Int. J. Mod. Phys. A {\bf 14}, 3597 (1999)
\bibitem{danielsson}
U.H. Danielsson, E. Keski-Vakkuri, M. Kruczeski, Nucl. Phys. B {\bf 563}, 279
(1999)
\bibitem{kalyana}
S. Kalyana Rama, B. Sathiapalan, Mod. Phys. Lett. A {\bf 14} 2635 (1999)
\bibitem{birmingham}
D. Birmingham, I. Sachs, S. N. Solodukhin, Phys. Rev. Lett. {\bf 88}, 151301
(2002)
\bibitem{son}
D. T. Son, A. O. Starinets, JHEP {\bf 0209}, 042 (2002)
\bibitem{horowitz}
G.T. Horowitz, and V. Hubeny, Phys. Rev. {\bf D62}, 0240027 (2000)
\bibitem{lemos}
V. Cardoso, J.P.S. Lemos, Phys. Rev. D {\bf 64}, 084017 (2001)
\bibitem{moss}
I.G. Moss, J.P. Norman, Class. Quant. Grav. {\bf 19}, 2323 (2002)
\bibitem{berti}
E. Berti, K.D. Kokkotas, Phys. Rev. {\bf D67}, 064020 (2003)
\bibitem{hod}
S. Hod, Phys. Rev. Lett. {\bf 81}, 4293 (2003)
\bibitem{bekenstein1}
J. Bekenstein, Lett. Nuovo Cimento {\bf 11}, 467 (1974)
\bibitem{bekenstein2}
J. Bekenstein, {\it Cosmology and Gravitation} edited by M. Novello
(Atlassciences, France 2000), pp. 1-85, gr-qc/9808028 (1998)
\bibitem{nollert}
H. P. Nollert, Phys. Rev D {\bf 47}, 5253 (1992)
\bibitem{natario}
V. Cardoso, J. Natario, R. Sciappa, hep-th/0403132 (2004)
\bibitem{WALD}
R. M. Wald, {\em General Relativity} (University of Chicago Press, Chicago,
1984)
\bibitem{gibbons}
G. W. Gibbons and S. W. Hawking, Phys. Rev. D {\bf 15} 2738n (1977)
\bibitem{mellor}
F. Mellor and I. G. Moss, Class. Quantum Grav. {\bf 6} 1379 (1989)
\bibitem{hawking}
S.W. Hawking, C.J. Hunter, M.M. Taylor-Robinson, Phys. Rev D {\bf 59}, 064005
(1999)
\bibitem{caldarelli}
M. M. Calarelli, G. Cognola and D. Klemm,
Class. Quant. Grav. {\bf 17}, 399 (2000)
\bibitem{gibbons04}
G. W. Gibbons, M. J. Perry and C. N. Pope, hep-th/0408217
\bibitem{ellis}
S. W. Hawking, G. F. R. Ellis, {\em The large scale structure of space-time},
(Cambridge University Press, 1973)
\bibitem{carter}
B. Carter, Commun. Math. Phys. {\bf 10}, 280 (1968)
\bibitem{demianski}
M. Demianski, Acta Astron. {\bf 23}, 211(1973)
\bibitem{chambers}
C.M. Chambers, I.G Moss, Class. Quantum Grav. {\bf 11}, 1035 (1994)
\bibitem{chandra}
S. Chandrasekhar, {\em The Mathematical Theory of Black Holes} (Claredon,
Oxford, 1983)
\bibitem{giammatteo}
M. Giammatteo, gr-qc/0303011 (2003)
\bibitem{sasaki}
Y. Mino, M. Sasaki, M. Shibata, H. Tagoshi, T. Tanaka, Prog. Theor. Phys. Suppl
{\bf 128}, (1997)
\bibitem{press}
W.H. Press, S.A. Teukoslky, B.F. Flannery, and W.T. Vetterling, {\it Numerical
Recipes, The Art of Scientific Computing}, (Cambridge University Press, 1986)
\bibitem{abramovitz}
M. Abramowitz, {\em Handbook of Mathematical Functions}, (Dover Publications,
inc., New York, 1965)
\bibitem{arfken}
G.B. Arfken, H.J. Weber, {\em Mathematical methods for physicists}, (Academic
Press, London 1995)
\bibitem{dias}
V. Cardoso, O.J.C. Dias, Phys Rev. D {\bf 70}, 084011 (2004)
\bibitem{konoplya}
V. Cardoso, R. Konoplya, J.P.S. Lemos, Phys. Rev. D {\bf 68}, 044024 (2003)
\bibitem{page}
S.W. Hawking, Don N. Page, Commun. Math. Phys. {\bf 87}, 577 (1983) 
\bibitem{banados1}
M. Banados, A. Gomberoff, C. Martinez, Class. Quant. Grav. {\bf 15}, 3575
(1998)
\bibitem{banados2}
M. Banados, Phys. Rev. D {\bf 57}, 1068 (1998)
\bibitem{hunter}
S.W Hawking, C.J. Hunter, M.M. Taylor-Robinson, Phys. Rev. D {\bf 59}, 064005
(1999)
\bibitem{reall}
S.W. Hawking, H.S. Reall, Phys. Rev D {\bf 61}, 0240014 (2000)
\bibitem{birmingham2}
D. Birmingham, Class. Quant. Grav. {\bf 16}, 1197 (1999)
\bibitem{johnson}
A.M. Awad, C.V. Johnson, Phys. Rev. D {\bf 63}, 124023 (2001)
\bibitem{awad}
A.M. Awad, Class. Quant. Grav. {\bf 20}, 2827 (2003)
\end{thebibliography}
\end{document}